\documentclass[superscriptaddress,amsmath,amssymb,aps,notitlepage,twocolumn,floatfix]{revtex4-1}

\usepackage{cmbright}
\DeclareFontShape{OT1}{cmss}{m}{it}{<->ssub*cmss/m/sl}{}

\newcommand{\uproman}[1]{\uppercase\expandafter{\romannumeral#1}}

\DeclareFontFamily{OT1}{cmbr}{\hyphenchar\font45 }
\DeclareFontShape{OT1}{cmbr}{m}{n}{%
  <-9>cmbr8
  <9-10>cmbr9
  <10-17>cmbr10
  <17->cmbr17
}{}
\DeclareFontShape{OT1}{cmbr}{m}{sl}{%
  <-9>cmbrsl8
  <9-10>cmbrsl9
  <10-17>cmbrsl10
  <17->cmbrsl17
}{}
\DeclareFontShape{OT1}{cmbr}{m}{it}{%
  <->ssub*cmbr/m/sl
}{}
\DeclareFontShape{OT1}{cmbr}{b}{n}{%
  <->ssub*cmbr/bx/n
}{}
\DeclareFontShape{OT1}{cmbr}{bx}{n}{%
  <->cmbrbx10
}{}

\usepackage[noindentafter]{titlesec}
\usepackage{titlesec}
\titleformat{name=\section}
  {\normalfont\large\bfseries\MakeUppercase}{\MakeUppercase{\thesection}}{0pt}{}
\titleformat{name=\subsection}
  {\normalfont\bfseries}{\thesection}{0pt}{}
\titlespacing{\section}{0cm}{0.7cm}{0.01cm}
\titlespacing{\subsection}{0cm}{0.45cm}{0cm}

\usepackage[utf8]{inputenc}
\usepackage{listings}
\usepackage{footmisc}
\usepackage{enumerate}
\usepackage{latexsym}
\usepackage{braket}
\usepackage{bm} 
\usepackage[version=4]{mhchem}
\usepackage{graphicx}
\usepackage[caption=false]{subfig}
\usepackage[colorlinks=True,linkcolor=red,citecolor=blue,urlcolor=blue]{hyperref}
\usepackage[capitalise]{cleveref}
\usepackage{appendix}
\usepackage{blkarray}
\usepackage{array}
\usepackage[dvipsnames]{xcolor}
\usepackage[normalem]{ulem}
\usepackage{wasysym}
\usepackage{multirow}
\usepackage{subfig}

\usepackage{caption}
\usepackage{lipsum}

\usepackage{xspace}

\usepackage{tabularx}
\usepackage{array}   
\newcolumntype{L}{>{$}l<{$}} 
\newcolumntype{R}{>{$}r<{$}} 
\newcolumntype{C}{>{$}c<{$}} 

\usepackage{float}

\usepackage[capitalise]{cleveref}
\usepackage{placeins}

\date{\today}

\usepackage{comment}

\begin{document}
    \title{
        Understanding the microscopic origin of the magnetic interactions in CoNb\textsubscript{2}O\textsubscript{6}}

\author{Amanda A. Konieczna}
\email{konieczna@itp.uni-frankfurt.de}
\affiliation{Institut f\"ur Theoretische Physik, Goethe-Universit\"at, 60438 Frankfurt am Main, Germany}
\author{David A. S. Kaib}
\email{kaib@itp.uni-frankfurt.de}
\affiliation{Institut f\"ur Theoretische Physik, Goethe-Universit\"at, 60438 Frankfurt am Main, Germany}

\author{Stephen M. Winter}
\email{winters@wfu.edu}
\affiliation{Department of Physics and Center for Functional Materials, Wake Forest University, Winston-Salem, North Carolina 27109, USA}

\author{Roser Valent\'i}
\email{valenti@itp.uni-frankfurt.de}
\affiliation{Institut f\"ur Theoretische Physik, Goethe-Universit\"at, 60438 Frankfurt am Main, Germany}

\date{\today}
	
\begin{abstract}
Motivated by the on-going discussion on the nature of magnetism in the quantum Ising chain CoNb$_2$O$_6$,  we present a first-principles-based analysis of its exchange interactions by applying an \textit{ab initio} approach with additional modelling that accounts for various drawbacks of a purely density functional theory ansatz. With this method we are able to extract and understand the origin of the magnetic couplings under inclusion of all symmetry-allowed terms, and to resolve the conflicting model descriptions
in CoNb$_2$O$_6$. We find that the twisted Kitaev chain and the transverse-field ferromagnetic Ising chain views are mutually compatible, although additional off-diagonal exchanges are necessary to provide a complete picture. We show that the dominant exchange interaction is a ligand-centered exchange process - involving the $e_g$ electrons -, which is rendered anisotropic by the low-symmetry crystal fields environments in CoNb$_2$O$_6$, giving rise to the dominant Ising exchange, while the smaller bond-dependent anisotropies  are found to originate from 
$d-d$ kinetic exchange
processes involving the $t_{2g}$ electrons. We demonstrate the validity of our approach by comparing the predictions of the obtained low-energy model to measured THz and inelastic neutron scattering spectra.
\end{abstract}

\maketitle
	
\section{INTRODUCTION}

\vspace{0.2cm}

For more than a decade, the quasi-one-dimensional Ising ferromagnet CoNb$_2$O$_6$ has been considered as a good experimental realization of the transverse-field ferromagnetic Ising chain (TFFIC)~\cite{maartense1977field-inducedmagtrans,scharf1979magneticstructures,hanawa1994anisotropicspecificheat,heid1995magphasediag,cabrera2014excitationsinthequantumparamagnetic,coldea2010quantumcriticalityinanisingchain,rutkevich2010ontheweakconfinement,kjaell2011boundstatesande8symmetry,nandi2019spinchargelattice,fava2020glidesymmetrybreaking,xu2022quantumcriticalmagneticexcitations,ringler2022singleionproperties}, showing the expected features of a transversal-field-induced quantum critical point at 5\,Tesla between a magnetically ordered and a quantum paramagnetic phase. At the critical field, the spin excitations change in character from domain-wall pairs in the ordered phase to spin-flips in the paramagnetic phase~\cite{coldea2010quantumcriticalityinanisingchain,fava2020glidesymmetrybreaking}.
Such a scenario has been confirmed by inelastic neutron scattering (INS)~\cite{coldea2010quantumcriticalityinanisingchain,coldea2023excitationsofquantumisingchain,coldea2023tuning}, specific heat~\cite{liang2015heatcapacitypeak}, nuclear magnetic resonance (NMR)~\cite{kinross2014evolutionofquantumfluctuations} and THz spectroscopy~\cite{morris2014hierarchyofboundstates,amelin2020experimentalobservation,armitage2021duality,amelin2022quantumspindynamics}
experiments. These early studies of CoNb$_2$O$_6$ were largely interpreted in terms of bond-independent XXZ-type anisotropic couplings, following expectations from classic works on Co$^{2+}$ magnetism \cite{lines1963magneticproperties,oguchi1965theoryofmagnetism}.

However, while the dominant physics of CoNb$_2$O$_6$ can be described by the TFFIC model, a (symmetry-allowed) staggered off-diagonal exchange contribution was proposed to be required to capture the details of the INS spectrum~\cite{fava2020glidesymmetrybreaking}, and a further refined TFFIC model was recently extracted from fitting to the INS spectrum at zero and high fields~\cite{coldea2023excitationsofquantumisingchain,coldea2023tuning}. A similar assertion was made on the basis of the field-evolution of the excitations in THz spectroscopy~\cite{armitage2021duality}, with the authors favoring an alternative twisted Kitaev chain model~\cite{you2014exact} to parameterize the bond-dependent interactions. The latter model was motivated, in part, by proposals of possible Kitaev-type bond-dependent couplings in high-spin Co$^{2+}$ compounds 
\cite{liu2018pseudospin,liu2020kitaevspinliquid,liu2021towardskitaevspinliquid,sano2018kitaevheisenberghamiltonian}, provided that a specific balance between competing exchange interactions is met. These proposals are in analogy with important developments in the last decades on the nature of exchange interactions in spin-orbit-coupled $4d$ and $5d$ transition-metal-based magnets in the context of bond-dependent Kitaev models~\cite{kitaev2006anyons,jackeli2009,haeyoung2014genericspinmodel,winter2016challenges,winter2017models,maksimov2019anisotropicexchange,takagi2019,trebst2022kitaev}. At present, the validity of this scenario for specific Co$^{2+}$ materials and the role of different exchange contributions remains a subject of discussion across multiple Co$^{2+}$ materials 
\cite{das2021xy,
chen2021spin,
sanders2022dominant,
winter2022magneticcouplings,
zhang2023magnetic,
liu2023non,
halloran2023geometrical,
xiang2023disorder,
bhattacharyya2024kitaev}. 
Adding to the discussion of CoNb$_2$O$_6$, further refinements of the model were recently suggested~\cite{churchill2024transforming} through a perturbation theory study and refitting of the experimental spectra. On the other hand,  based on symmetry arguments, the authors of Ref.~\cite{gallegos2024MagnonInteractionsQuantum} provided valuable insights into the general structure of the spin-anisotropic model of CoNb$_2$O$_6$ and studied effects of magnon interactions in its excitation spectrum by adopting the model parameters from Ref.~\cite{coldea2023excitationsofquantumisingchain,coldea2023tuning}.

\begin{figure}
   \begin{minipage}[b]{\linewidth}
    \includegraphics[width=.85\linewidth]{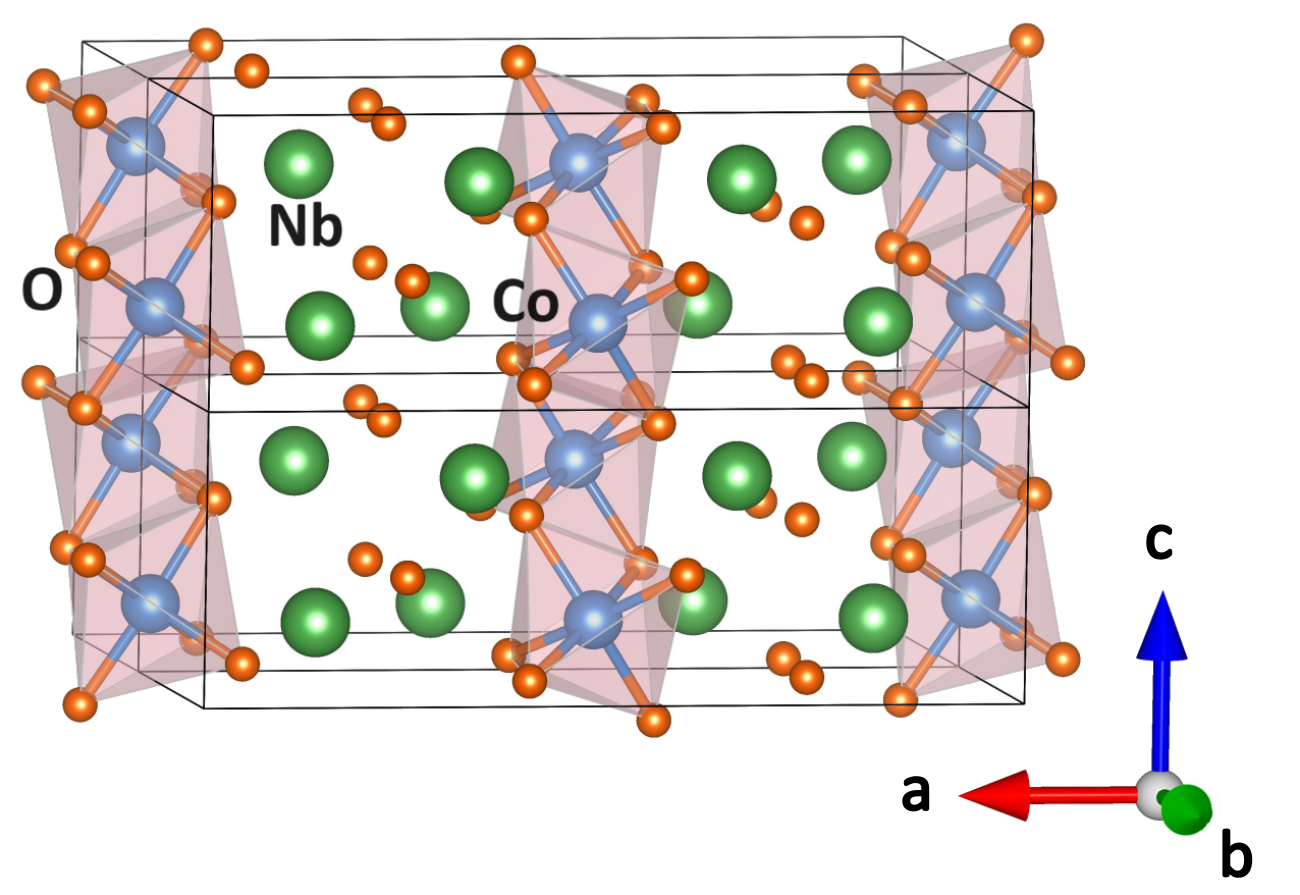}
    \caption{CoNb$_2$O$_6$ structure with Co$^{2+}$ zigzag chains expanding along the $c$-axis. The octahedral oxygen environment of the Co-ions is visible. The Co$^{2+}$  chains align antiferromagnetically to each other forming an antiferromagnetic system.}
    \label{fig:structure}
   \end{minipage}
\end{figure}

In reality, the various experimentally motivated parametrizations of the couplings in CoNb$_2$O$_6$ lead to similar Hamiltonians, but the apparently conflicting interpretations call for a detailed microscopic analysis of the exchange contributions.

In this work we present a detailed study of the exchange interactions in CoNb$_2$O$_6$, obtained by applying an \textit{ab-initio} approach with additional modelling that accounts for various drawbacks of a purely density functional theory (DFT)-type ansatz. The method allows us to extract and understand the origin of the magnetic couplings under inclusion of all symmetry-allowed terms, and to resolve the conflicting model descriptions above. We find that the twisted Kitaev chain and TFFIC views 
are microscopically consistent only if the dominant exchange interaction is a ligand-centered exchange process, which is rendered anisotropic by the low-symmetry crystal fields environments in CoNb$_2$O$_6$, giving rise to the dominant Ising exchange. The smaller bond-dependent anisotropies are found to arise from a combination of terms we discuss in detail. We demonstrate the validity of our approach by comparing the predictions of the obtained low-energy model to measured THz and INS spectra. Finally, this model is compared to previously proposed models in the perspectives of both TFFIC and twisted Kitaev chain descriptions.

\begin{figure}
   \begin{minipage}[b]{\linewidth}
    \includegraphics[width=.9\linewidth]{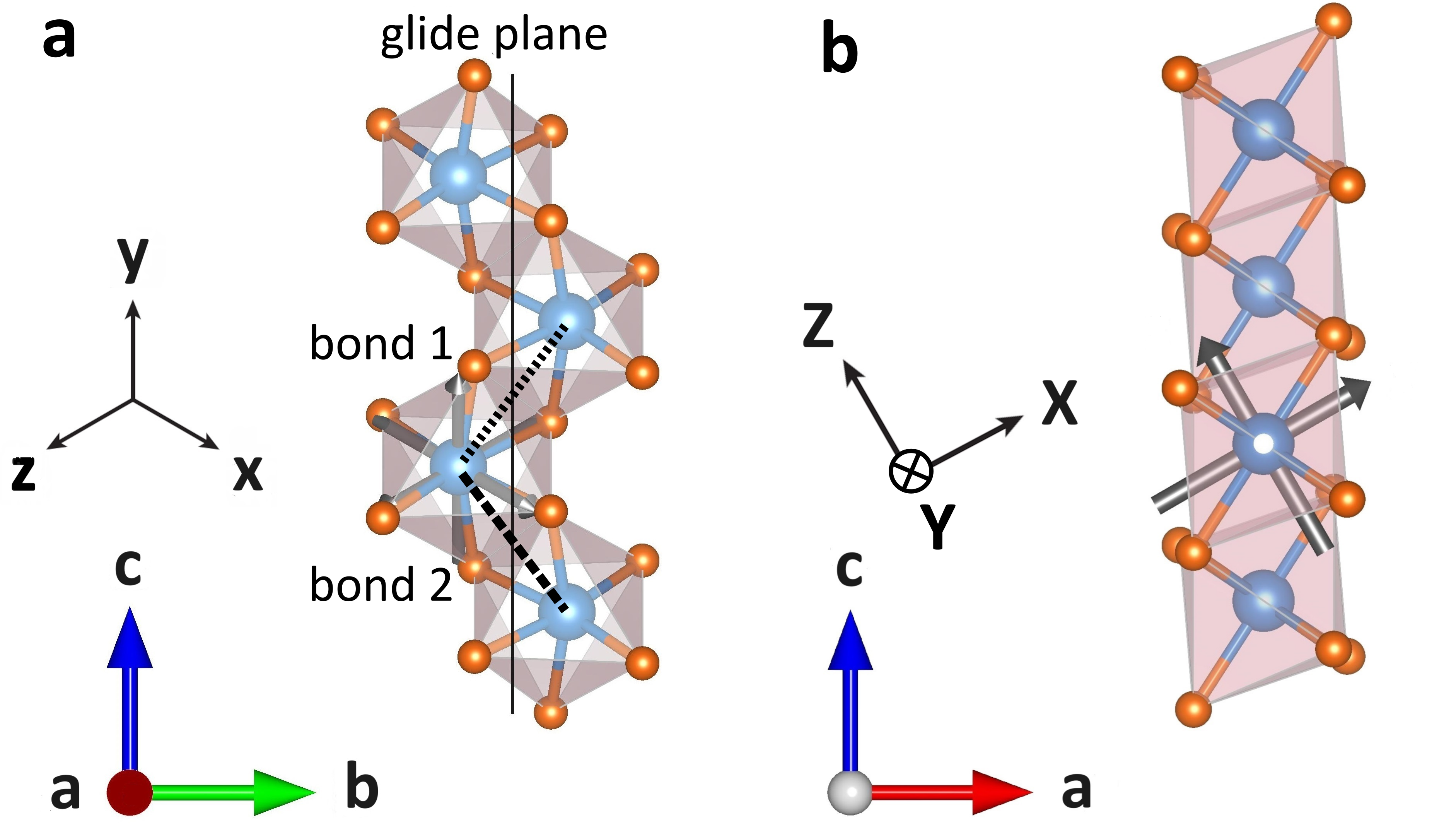}
    \caption{Different coordinate systems shown in a Co-chain with surrounding oxygen octahedra. Due to the antiferromagnetic alignment of the chains the local coordinate systems depend on the choice of the chain. \textbf{a} Cubic axes $\textbf{x}, \textbf{y}, \textbf{z}$. They correspond to the conventional Kitaev coordinates. The glide plane perpendicular to the crystallographic \textbf{b}-axis is visible, as well as the two bonds defined in the twisted Kitaev notation. {\it Bond 1} is perpendicular to the \textbf{x}-axis, 
    {\it bond 2} to the \textbf{z}-axis. \textbf{b} Principal axes $\textbf{X}, \textbf{Y}, \textbf{Z}$ as described in Eq.~\ref{eq:coordinates_XYZ} for $\phi=30^\circ$.}
    \label{fig:methods:structure_big_axes}
   \end{minipage}
\end{figure}

\vspace{0.2cm}
\section{FIRST PRINCIPLES CALCULATIONS}
\vspace{0.2cm}
\subsection{Structure and Symmetries}

\vspace{0.2cm}

In the following sections, all calculations are based on the crystal structure reported in Ref.~\cite{heid1995magphasediag}. CoNb$_2$O$_6$ crystallizes in the orthorhombic space group \textit{Pbcn} featuring pseudospin 1/2 Co$^{2+}$ zigzag chains extending along the crystallographic c-axis (Fig.~\ref{fig:structure}). The magnetic Co$^{2+}$ ions are linked through edge-sharing oxygen octahedra, forming chains characterized by alternating bonds. The ferromagnetic alignment of Co$^{2+}$ spins within each chain occurs at temperatures below 1.97K. The octahedral environments in adjacent chains are crystallographically equivalent, though their opposing alignment results in a weak antiferromagnetic exchange between the chains, as well as in the presence of two distinct easy axes \cite{heid1995magphasediag,kohn1999threedimensional,wanklyn1979magneticstructures,SetsuoMitsuda1994magordering}.

In the discussion of the electronic Hamiltonian and magnetic couplings, it is useful to refer to two sets of coordinate systems. The ``cubic'' axes are defined as an orthonormal basis of vectors \textbf{x}, \textbf{y} and \textbf{z}, which point approximately along the octahedral Co-O bonds, as shown in Fig.~\ref{fig:methods:structure_big_axes} {\bf a}. In terms of the global crystallographic axes (\textbf{a}, \textbf{b}, \textbf{c}), these are:
\begin{align}
    \begin{aligned}
         \textbf{x} = \left(\begin{array}{c} 1/\sqrt3 \\ 1/\sqrt2 \\ -1/\sqrt6 \end{array}\right),\ \textbf{y} = \left(\begin{array}{c} 1/\sqrt3 \\ 0 \\ 2/\sqrt6 \end{array}\right),\ \textbf{z} = \left(\begin{array}{c} 1/\sqrt3 \\ -1/\sqrt2 \\ -1/\sqrt6 \end{array}\right)
    \end{aligned}
\end{align}
In this coordinate system, the \textbf{y}-axis lies within the \textbf{c}-glide plane, which lies halfway between the \textbf{x}- and \textbf{z}-axes.

As is visible in Fig.~\ref{fig:methods:structure_big_axes} {\bf a}, one of the cubic axes is situated within the a-c plane with an angle of around $\pm30^{\circ}$ from the {\bf c}-axis, where the sign of the angle alternates from chain to chain. Initial studies assumed that each easy axis would align with this cubic axis, leading to some disagreement in the literature (see \cite{heid1995magphasediag,kohn1999threedimensional,wanklyn1979magneticstructures,SetsuoMitsuda1994magordering}) on that matter. As discussed below in further detail, we find that the direction of each easy axis lies in opposition to the corresponding cubic axis, meaning that for an angle of $\pm30^{\circ}$ for the cubic axis, the easy axis lies at $\mp30^{\circ}$ from the {\bf c}-axis (see Fig.~\ref{fig:methods:structure_big_axes}).

In order to enable comparisons between our results and those obtained experimentally in Refs.~\cite{coldea2023tuning,armitage2021duality}, we also refer to a coordinate system of ``principal'' axes denoted by uppercase letters \textbf{X}, \textbf{Y} and \textbf{Z}, where \textbf{Z} is aligned with the easy axis, as depicted in Fig.~\ref{fig:methods:structure_big_axes} {\bf b}. With respect to (\textbf{a}, \textbf{b}, \textbf{c}), these ``principal'' axes are defined as:
\begin{align}
    \begin{aligned}
        \textbf{X} = \left(\begin{array}{c} \cos\phi \\ 0 \\ \sin\phi \end{array}\right),\ \textbf{Y} = \left(\begin{array}{c} 0 \\ 1 \\ 0 \end{array}\right),\ \textbf{Z} = \left(\begin{array}{c} -\sin\phi \\ 0 \\ \cos\phi \end{array}\right)
    \end{aligned}
    \label{eq:coordinates_XYZ}
\end{align}
where $\phi=+30^\circ$ is the canting angle. This definition sets the local \textbf{Z}-axis to be tilted by $\phi$ from the crystallographic $c$-axis in the negative $a$-direction. This choice of direction for \textbf{Z} is based on our estimates on the Ising axis orientation from determining the g-tensor. In this coordinate system, the $\mathbf Y$-axis is perpendicular to the $c$-glide plane, while the \textbf{X}- and \textbf{Z}-axes lie within it.

\vspace{0.2cm}
\subsection{Multi-Orbital Hubbard Model}

\vspace{0.2cm}

In order to estimate the magnetic couplings, we apply an {\it ab-initio} based approach~\cite{riedl2019} (projED), which has been shown to yield reliable results for highly anisotropic spin Hamiltonians~\cite{winter2017breakdown,winter2018probing,riedl2022microscopicorigin,razpopov2023j}. The method is based on two steps. First, an electronic multi-orbital Hubbard Hamiltonian is obtained from a full-relativistic DFT calculation on the basis of appropriately constructed Wannier orbitals, as well as suitably defined Coulomb interactions. Second, the multi-orbital Hubbard Hamiltonian is exactly diagonalized on finite-size clusters. The eigenstates are then projected into the low-energy space and mapped to an effective spin Hamiltonian. For this purpose, we employ ideal $j_{1/2}$ states described in \nameref{appendix:groundstateprojection}.

We start with the description of all necessary terms contributing to the multi-orbital Hubbard Hamiltonian. The Hamiltonian in the basis of Co 3$d$-orbitals is given by:
\begin{align}\label{eq:methods:general_hamiltonian}
    \mathcal H = \mathcal H_{\rm hop} + \mathcal H_{\rm nn-U} + \sum_i \mathcal H_i
\end{align}
The first term in Eq.~\eqref{eq:methods:general_hamiltonian} describes the hoppings between orbitals $a$ and $b$ on two Co-sites $i$ and $j$: 
\begin{align}\label{eq:methods:hopping_hamiltonian}
    \mathcal H_{\rm hop} = \sum_{\sigma\sigma'}\sum_{ab} t_{ij,\sigma\sigma'}^{ab}d_{ia\sigma}^{\dagger}d_{jb\sigma'}
\end{align}
$\mathcal{H}_\text{nn-U}$ describes the Coulomb interaction between different Co-ions $i$ and $j$ through the ligands. This is given by:
\begin{align}\label{eq:methods:nnHamiltonian_short}
    \begin{aligned}
        \mathcal H_{\rm nn-U} = \sum_{ab} \Bigg[ & \tilde U_1 \left(n_{ia\uparrow}n_{jb\downarrow}+n_{ja\uparrow}n_{ib\downarrow}\right) + 2\tilde U_2 \sum_{\sigma} n_{ia\sigma}n_{jb\sigma} \\
        & + \tilde J \left( d_{ia\uparrow}^{\dagger}d_{jb\downarrow}^{\dagger}d_{ia\downarrow}d_{jb\uparrow} + d_{ja\uparrow}^{\dagger}d_{ib\downarrow}^{\dagger}d_{ja\downarrow}d_{ib\uparrow} \right) \Bigg],
    \end{aligned}
\end{align}
which resembles the typical intersite Coulomb term consisting of diagonal density-density contributions as well as a ``Hund's''-type coupling. Explicit inclusion of this latter term to the electronic Hamiltonian is crucial in order to account for ligand-centered exchange processes downfolded into the Co $d$-orbital basis. Especially the $e_g$-orbitals hybridize strongly with the ligands and produce large contributions to this term, which are manifestly important for correctly describing the magnetism of edge-sharing $3d$ materials (see, for example, \cite{autieri2022limited}). The terms $\tilde U_1$, $\tilde U_2$ and $\tilde J$ as well as the details of the derivation are discussed in the {\it Intersite Coulomb Interactions} section. 

Finally, the last term in \cref{eq:methods:general_hamiltonian} includes all relevant on-site effects for each site $i$:
\begin{align}\label{eq:methods:onsite_hamiltonian}
    \mathcal H_i = \mathcal H_{\rm CF} + \mathcal H_{\rm SOC} + \mathcal H_{\rm U}
\end{align}
First, $\mathcal H_{\rm CF}$ accounts for the crystal field splitting and $\mathcal H_{\rm SOC}$ refers to spin-orbit coupling. The interaction term $\mathcal H_{\rm U}$ includes all on-site contributions to the Coulomb interaction, which in its most general form is:
\begin{align}
    \mathcal H_{\rm U} = \frac{1}{2} \sum_{\sigma,\sigma'}\sum_{a,b,c,d} U_{abcd}d_{ia\sigma}^{\dagger}d_{ib\sigma'}^{\dagger}d_{ic\sigma'}d_{id\sigma}
\end{align}
Here $a$, $b$, $c$ and $d$ label the $d$-orbitals while $\sigma$ and $\sigma'$ define the spin orientation. The $U$-matrix-elements are considered within the spherically symmetric approximation and are fully determined by the Slater integrals $F_0$, $F_2$ and $F_4$ via
\begin{align}
    U_{avg} = & F_0 + \frac{8}{7}J_{avg}  & \text{and}  &   &   J_{avg} = & \frac{F_2 + F_4}{14}.
\end{align}
Following Ref.~\cite{Pavarini_2014}, the ratio $F_4=\frac{5}{8}F_2$ is applied, which is a frequently used assumption for $3d$ orbitals. For the computations, the parameters $U_{avg}=5.1$eV and $J_{avg}=0.9$eV will be used, which have been obtained from constrained DFT calculations for CoO in Ref.~\cite{scheffler2010firstprinciplesmodeling}.

The material-specific contributions to the Hamiltonian contained in $\mathcal{H}_\text{hop}$, $\mathcal{H}_{\rm CF}$, $\mathcal H_{\rm SOC}$, and $\mathcal H_{\rm nn-U}$ were estimated on the basis of DFT calculations. To that end, for $\mathcal H_{\rm hop}$, and $\mathcal H_{\rm CF}+\mathcal H_{\rm SOC}$ we perform a non-magnetic fully relativistic FPLO (see Ref.~\cite{FPLO1999}) calculation on a 12x12x12 grid within the generalized gradient approximation (GGA). Projective Wannier functions were used to obtain hopping parameters. The nearest-neighbor spin-diagonal hoppings can be found in \nameref{appendix:hopping_NNCoulomb_parameters}. The parametrization of the intersite Coulomb interactions is discussed below. We first refine the crystal field in order to ensure consistency with the measured $g$-tensors.

\vspace{0.2cm}
\subsection{Crystal Field Splitting and $g$-Tensor}

\vspace{0.2cm} 

In this section, we first discuss the modelling of the local crystal field, and its consequences on the $g$-tensor.

In Co $3d^7$ compounds, the relative weakness of the SOC effectively enhances the sensitivity of the low-energy spin interactions and $g$-tensor to crystal-field distortions, which influence the specific spin-orbital composition of the local moments. As a consequence, accurate modelling of the local single-ion state is a prerequisite for understanding the intersite exchange couplings. As demonstrated in Ref.~\onlinecite{mou2024comparative}, Wannier fitting only sometimes achieves the necessary accuracy to reproduce the experimental crystal field energies. We therefore introduce some parameter adjustments in our electronic Hamiltonian to ensure consistency with the experimentally determined $g$-tensor. Experimentally, the $g$-tensor has been estimated in various works. Ref.~\cite{ringler2022singleionproperties} fitted $g$-tensor principal values and orientations from EPR and INS spectra for two different samples. They find two of the $g$-values to be relatively close to another, and a much larger third $g$-value, where the principal axes are in agreement with the definition in Eq.~\eqref{eq:coordinates_XYZ} for an angle $\phi=37^\circ$. Ref.~\cite{coldea2023tuning} presents a similar picture, with the principal values presented in Table~\ref{tab:results:gtensor} for ``Fit$^{\rm INS}$''. There, the principal axes align with Eq.~\eqref{eq:coordinates_XYZ} for $\phi=30^\circ$.

In order to compute the $g$-tensor from first principles, we diagonalize the on-site Hamiltonian Eq.~\eqref{eq:methods:onsite_hamiltonian} with an additional term $\mathcal{H}_{\rm Zeeman}$ that describes the coupling
of the angular momentum $\vec L$ and spin $\vec S$ of the Co-ions to an external magnetic field $\vec B$:
\begin{align}\label{eq:methods:magHam}
    \mathcal H = \mathcal H_i + \mathcal H_{\rm Zeeman} = \mathcal H_i + \mu_B \vec B \cdot \left( \vec L_i + 2\vec S_i \right).
\end{align}
After projecting the resulting low-energy ($2\times 2$) Hamiltonian onto ideal $j_{1/2}$ states, the $g$-tensor is obtained by taking numerical derivatives with respect to different components of $\vec B$. Due to a 2-fold rotation axis parallel to the crystallographic $b$-axis going through each Co site, one of the principal axes of the $g$-tensor is enforced to be parallel to the {\bf b}-axis (coincident with the {\bf Y}-axis). 

As a starting point for discussion, we first consider the crystal field splitting matrix $\mathbb{M}_\textbf{CF}$ obtained directly from the DFT calculation after Wannierization. Written in terms of orbitals $d_{xy}$, $d_{yz}$, $d_{xz}$, $d_{z^2}$ and $d_{x^2-y^2}$, this is (in meV):
\begin{align}
    \mathbb{M}_{\textbf{CF}}^{\rm DFT} = \begin{pmatrix}
    0 & -8.7 & 14.5 & 16.3 & 29.9 \\
    -8.7 & 0 & 14.5 & 17.7 & 29.1 \\
    14.5 & 14.5 & 89.9 & 12.2  & 21.2 \\
    16.3 & 17.7 & 12.2 & 915.8 & -9.4 \\
    29.9 & 29.1 & 21.2 & -9.4 & 905.1 \\
    \end{pmatrix}
    \label{eq:methods:CFS_matrix}
\end{align}
where for readability these values were gauged such that $\mathbb{M}_{xy,xy}=\mathbb{M}_{yz,yz}=0$. Performing the calculation for the $g$-tensor starting from $\mathbb{M}_\textbf{CF}^{\rm DFT}$ yields principal axes of the $g$-tensor corresponding to \textbf{X}, \textbf{Y}, \textbf{Z} of \cref{eq:coordinates_XYZ}, but with an incorrect canting angle of $\phi\approx 45^\circ$ (complete $g$-tensor given in \nameref{appendix:detailed_results}), instead of $\phi \approx 30^\circ$ from experiment \cite{heid1995magphasediag,coldea2023tuning}. Since the magnetic couplings depend strongly on the crystal field, $\mathbb{M}_\textbf{CF}$ does not represent an adequate starting point for further calculations. 

After investigating the effects of various alterations to $\mathbb{M}_\textbf{CF}^{\rm DFT}$, we conclude that this discrepancy is likely due to an overestimation of the splitting between the $xy$/$yz$-orbitals and the $xz$-orbital. Such a discrepancy can arise in various ways. First, we rely on the accuracy of the starting structure; small changes in the atomic positions can lead to significant effects on the computed CFS matrix. Second, inherent to the Wannier interpolation in DFT is an effective inclusion of the mean-field Coulomb terms into the single-particle Hamiltonian. In relatively undistorted local environments, these mean-field contributions to the orbital energies can be expected to be relatively orbital-independent within the $d$-orbitals, i.e. provide only a constant diagonal shift of the orbital energies. However, with lower symmetry environments, the mean-field contributions may instead display significant orbital-dependence. It is therefore more appropriate to improve the crystal field estimates by fitting to available experimental data, whenever possible.

In the present case, we find that the minimal modification of $\mathbb{M}_\textbf{CF}^{\rm DFT}$ required to reflect the major aspects of the experimental $g$-tensor (see Refs. \cite{ringler2022singleionproperties,heid1995magphasediag}) is simply to set $\mathbb{M}_{xz,xz}$ in Eq.~\eqref{eq:methods:CFS_matrix} to 15meV, i.e.\ reducing the $t_{2g}$ splitting. This results in principal axes of the $g$-tensor that correspond to a canting of $\phi\approx 30.8^\circ$ in \cref{eq:coordinates_XYZ}, and yields the principal $g$-values shown in Table~\ref{tab:results:gtensor} under ``Model$_{\rm corr}^{\rm DFT}$'' (complete $g$-tensor is given in \nameref{appendix:detailed_results}). The modification of the crystal field reduces the previously large difference between $g_X$ and $g_Y$ (see \nameref{appendix:detailed_results}), and reproduces the large anisotropy of the $g_X/g_Y$ and $g_Z$ principal values. Altogether, the comparison to experimentally fitted $g$-values of Ref.~\cite{coldea2023tuning} (``Fit$^\mathrm{INS}$'' in Table~\ref{tab:results:gtensor}) yields a very good agreement.

\begin{table}[t]
    \centering
    \begin{tabular}{|c|c|c|}
        \cline{2-3}
        \multicolumn{1}{c|}{} & \multicolumn{1}{c|}{}  & \multicolumn{1}{c|}{} \\[-3pt]
        \multicolumn{1}{c|}{} &  $\ \ $Fit$^\mathrm{INS}$$\ \ $ &  $\ $Model$_\mathrm{corr}^\mathrm{DFT}$$\ $ \\[5pt]
        \hline
        \multicolumn{1}{|c|}{} & \multicolumn{1}{c|}{}  & \multicolumn{1}{c|}{} \\[-3pt]
        $\quad g_X\quad$ & 3.29(6)  & 2.98 \\[5pt]
        $g_Y$ & 3.32(2)  & 3.13 \\[5pt]
        $g_Z$ & 6.90(5)  & 6.90 \\[5pt]
        \hline
    \end{tabular}
    \caption{g-tensor principal values in \textbf{XYZ} coordinates (\cref{eq:coordinates_XYZ}, $\phi\approx 30^\circ$) from fitting to INS data
    (Fit$^{\rm INS}$)~\cite{coldea2023tuning} next to our results when considering the corrected crystal field matrix (Model$_\mathrm{corr}^{\rm DFT}$). }
    \label{tab:results:gtensor}
\end{table}

From now on, we perform all calculations of the intersite couplings employing Model$_\mathrm{corr}^{\rm DFT}$ that includes the corrected crystal field splitting, setting $\mathbb{M}_{xz,xz}$ in Eq.~\eqref{eq:methods:CFS_matrix} to 15meV.

\vspace{0.2cm}

\subsection{Intersite Coulomb Interactions}

\begin{figure}
   \begin{minipage}[b]{\linewidth}
    \includegraphics[width=.8\linewidth]{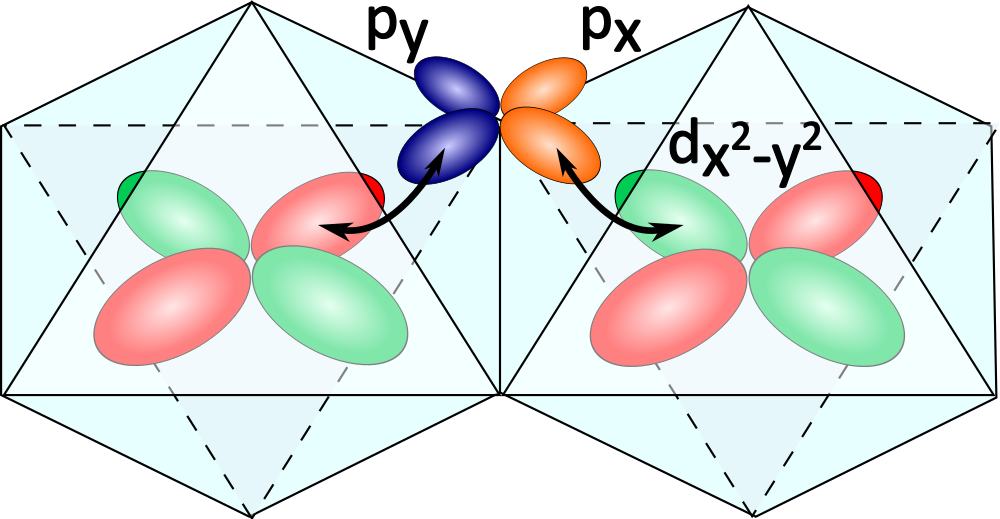}
    \caption{Two neighboring octahedra with Co-atoms at each center. The hybridization of their $e_g$-orbitals with ligand $p$-orbitals leads to an effective inter-site Coulomb repulsion when holes interacting at the ligand repel each other. Here the process is visualized for $d_{x^2-y^2}$ orbitals, in which the hybridization is dominant.}
    \label{fig:test}
   \end{minipage}
\end{figure}

The nearest-neighbor magnetic interactions between edge-sharing octahedra of high-spin Co$^{2+}$ ions are strongly influenced by ferromagnetic Coulomb exchange interactions through the ligands \cite{winter2022magneticcouplings}, similar as for other $3d$ ions \cite{autieri2022limited}. In the standard picture of these contributions considering explicitly the metal $d$ and ligand $p$ orbitals, the ferromagnetic contribution arises at order $J_H^p t_{pd}^4/\Delta_{pd}^4$, where $J_H^p$ refers to the Hund's coupling of the ligands, $t_{pd}$ is the hopping between the $p$ and $d$ orbitals, and $\Delta_{pd}$ is the charge transfer energy. These are discussed as ``charge-transfer'' processes in Ref.~\cite{liu2018pseudospin}. They capture the effect that two holes, nominally associated with different sites, may meet on the same ligand, and experience Hund's coupling that reduces the energy of the triplet configuration (see Fig.~\ref{fig:test}). When projected into the basis of $d$ Wannier functions, this term is reinterpreted as an effective intersite Coulomb interaction. The weight of the $p$-orbitals in these Wannier functions scale like $(t_{pd}/\Delta_{pd})^2$. As a consequence of the fact that the $d$-orbital Wannier functions of different Co sites have finite coefficients on the same ligand, rotation of the ligand Coulomb interactions into the Wannier basis results in intersite Coulomb interactions of order $J_H^p t_{pd}^4/\Delta_{pd}^4$. 

To make this qualitative discussion concrete and to arrive at the interaction $\mathcal{H}_\text{nn-U}$ summarized in \cref{eq:methods:nnHamiltonian_short}, one can write the creation operators associated with the bare $p$-orbitals $\alpha$ on ligand atom $n$ in terms of $p$-orbital Wannier functions located at the same site, and $d$-orbital Wannier functions $a$ on Co site $i$ through:
\begin{align}\label{eq:methods:orbitalansatz}
    p_{n\alpha\sigma}^{\dagger} = \phi_{n\alpha\sigma}\tilde{p}_{n\alpha\sigma}^\dagger + \sum_{i}\sum_{a} \phi_{ia}^{n\alpha}\tilde{d}_{ia\sigma}^{\dagger}. 
\end{align}
where the tilde indicates mixed $p/d$ Wannier orbitals. In the downfolded basis, only the $\tilde{d}$ Wannier orbitals are explicitly considered. The bare Coulomb interaction for the $p$-orbitals, as it is given for $p$-orbitals in \cite{spencer2016hubbardlikehamiltonians}, is:
\begin{align}\label{eq:methods:pCoulomb}
    \mathcal H = \frac{1}{2} \sum_{n}\sum_{\sigma\sigma'}\sum_{\alpha\beta\gamma\delta} V_{\alpha\beta\gamma\delta} p_{n\alpha\sigma}^{\dagger}p_{n\beta\sigma'}^{\dagger}p_{n\delta\sigma'}p_{n\gamma\sigma}
\end{align}
with
\begin{align}
    V_{\alpha\beta\gamma\delta} = U\delta_{\alpha\gamma}\delta_{\beta\delta} + J_H\left(\delta_{\alpha\delta}\delta_{\beta\gamma}+\delta_{\alpha\beta}\delta_{\gamma\delta}\right)
\end{align}
Inserting the ansatz in Eq.~\eqref{eq:methods:orbitalansatz} into Eq.~\eqref{eq:methods:pCoulomb}, and retaining those terms with all $d$-orbital operators leads to the expression shown in Eq.~\eqref{eq:methods:nnHamiltonian_short}, for which the coupling matrices $\tilde U_1$, $\tilde U_2$ and $\tilde J$ are defined as:
\begin{subequations}\label{eq:methods:nnHunds_prefacs}
    \begin{align}
        \tilde U_1 = & \left(U_0-2J\right) A_{ab} + 2JB_{ab} + U_0C_{ab} \\
        \tilde U_2 = & \frac{U_0-3J}{2}  \left(A_{ab} - B_{ab}\right) \\
        \tilde J = & J_HA_{ab} + (U_0-J_H) B_{ab} + U_0C_{ab},
    \end{align}
\end{subequations}
Here, $A_{ab} \equiv \gamma\sum_{n,\alpha\neq\beta} |\phi_{ia}^{n\alpha}|^2 |\phi_{jb}^{n\beta}|^2$, $B_{ab} \equiv  \gamma\sum_{n,\alpha\neq\beta} \phi_{ia}^{n\alpha}\phi_{ia}^{n\beta}\phi_{jb}^{n\alpha}\phi_{jb}^{n\beta}$ and $C_{ab} \equiv \gamma\sum_{n,\alpha} |\phi_{ia}^{n\alpha}|^2 |\phi_{jb}^{n\alpha}|^2$.

\vspace{0.2cm}
\noindent {$J_H=\frac{4}{25}F_2^p$} refers to the Hund's exchange energy associated with the O~2$p$ orbitals and $F_2^p$ to the according Slater integral. Following Ref.~\cite{johnson1990electronicstructureofAg2O}, $F_2^p$ for oxygen $p$-orbitals is expected to lie at 6eV. $U_0$ corresponds to $F_0^p+\frac{4}{3}J$. For our calculations we use $J=0.3U_0$ in analogy to Ref.~\cite{liu2018pseudospin}, and estimate $A_{ab}, B_{ab},$ and $C_{ab}$ from non-relativistic FPLO calculations. Finally, we introduced a scaling factor $\gamma$ to the overlap integrals, in order to account for screening of the Coulomb interactions with respect to the bare Slater integral.

The nearest-neighbor Coulomb interactions influence the magnetic exchange primarily through the $\tilde J$-terms, which result in a ferromagnetic interaction when projected into the $j_{1/2}$ basis. The $\tilde U_1$ and $\tilde U_2$ represent spin-independent density-density interactions, and have little influence on the magnetic couplings (see \nameref{appendix:detailed_results} for comparison). The obtained $\tilde J$ matrix on {\it bond 2} (see Fig.~\ref{fig:structure} {\bf b}), in terms of $d_{xy}$, $d_{yz}$, $d_{xz}$, $d_{z^2}$ and $d_{x^2-y^2}$, is ($\tilde U_1$ and $\tilde U_2$ are given in \nameref{appendix:hopping_NNCoulomb_parameters}):
\begin{align}
    \begin{aligned}
        \tilde J &= J_HA_{ab} +  (U_0-J_H) B_{ab} + U_0C_{ab}
        \\ &= \ \begin{pmatrix}
        0.31 & 0.18 & 0.26 & 0.74  & 2.45 \\
        0.18 & 0 & 0.58 & 0.18  & 0.60 \\
        0.26 & 0.58 & 0 & 0.16  & 0.44 \\
        0.74 & 0.18 & 0.16 & 0.24  & 0.79 \\
        2.45 & 0.60 & 0.44 & 0.79  & 2.54 \\
        \end{pmatrix}\,\mathrm{meV}
    \end{aligned}
\end{align}
The dominant contributions to intersite coupling (in all of $\tilde U_1$, $\tilde U_2$, $\tilde J$) stem from the $e_g$-orbitals, due to their much larger hybridization with the ligand $p$-orbitals. As discussed in \cite{liu2018pseudospin,sano2018kitaevheisenberghamiltonian,winter2022magneticcouplings}, it is the presence of the $e_g$ electrons in the high-spin $d^7$ configuration that leads to a significant effect of the ferromagnetic ligand exchange processes.

\section{RESULTS}

\subsection{Magnetic Exchange Couplings}
\vspace{0.2cm}

As introduced above, CoNb$_2$O$_6$ has been discussed both, in terms of a twisted Kitaev chain as well as an Ising model supplemented by isotropic and anisotropic couplings \cite{armitage2021duality,coldea2010quantumcriticalityinanisingchain,coldea2023excitationsofquantumisingchain,coldea2023tuning,churchill2024transforming,gallegos2024MagnonInteractionsQuantum}. The former perspective is made apparent in the cubic {\bf{xyz}} coordinate system of Fig.~\ref{fig:methods:structure_big_axes} {\bf a}, while the latter is natural in the principal {\bf{XYZ}}-coordinates of Fig.~\ref{fig:methods:structure_big_axes} {\bf b}.

Following \cref{fig:methods:structure_big_axes} {\bf a}, {\it bond 1} is defined as perpendicular to the cubic {\bf x}-direction and {\it bond 2} perpendicular to {\bf z}. The symmetry-allowed magnetic exchange on {\it bond 1} can then be cast into the form
\begin{align}
    \begin{aligned}
        \mathcal H_{\rm eff}^{\rm bond 1} = & J_K\Vec{S}_i\Vec{S}_{i+1} + \bar{K}S_i^yS_{i+1}^y + KS_i^zS_{i+1}^z \\
        & + \Gamma_1 \left(S_i^xS_{i+1}^y+S_i^yS_{i+1}^x\right) \\
        & + \Gamma_2\left(S_i^xS_{i+1}^z+S_i^zS_{i+1}^x\right) \\
        & + \Gamma_3\left(S_i^yS_{i+1}^z+S_i^zS_{i+1}^y\right)
    \end{aligned}
    \label{eq:results:mag_ham1_kitaev}
\end{align}
in cubic {\bf xyz}-coordinates. Here, $K$ is the Kitaev coupling, $J_K$ is conventional Heisenberg exchange, and $\Gamma_i$ are off-diagonal symmetric couplings. The exchanges on both bonds are related through a glide-plane symmetry:
\begin{align}
    \begin{aligned}
        &S_i^x \Rightarrow -S_{i+1}^z \\
        &S_i^y \Rightarrow -S_{i+1}^y \\
        &S_i^z \Rightarrow -S_{i+1}^x, \\
    \end{aligned}
\end{align}
which implies for the second bond:
\begin{align}
    \begin{aligned}
        \mathcal H_{\rm eff}^{\rm bond 2}
        = & J_K\Vec{S}_i\Vec{S}_{i+1} + KS_i^xS_{i+1}^x + \bar{K}S_i^yS_{i+1}^y \\
        & + \Gamma_3 \left(S_i^xS_{i+1}^y+S_i^yS_{i+1}^x\right) \\
        & + \Gamma_2\left(S_i^xS_{i+1}^z+S_i^zS_{i+1}^x\right) \\
        & + \Gamma_1\left(S_i^yS_{i+1}^z+S_i^zS_{i+1}^y\right).
    \end{aligned}
    \label{eq:Hspin2}
\end{align}

An alternative parameterization of the Hamiltonian is given in terms of the principal {\bf XYZ}-axes of \cref{fig:methods:structure_big_axes} {\bf b} and \cref{eq:coordinates_XYZ} with $\phi=30^\circ$:  
\begin{align}
    \begin{aligned}
        \mathcal H_{\rm eff} = - J \sum_i & \Big\{ S_i^ZS_{i+1}^Z + \mu_YS_i^YS_{i+1}^Y + \mu_XS_i^XS_{i+1}^X \\
        & + \mu_{XZ} \left(S_i^XS_{i+1}^Z + S_i^ZS_{i+1}^X\right) \\
        & + (-1)^i \Bigl[ \mu_{XY} \left(S_i^XS_{i+1}^Y + S_i^YS_{i+1}^X\right) \\
        & \qquad \quad + \mu_{YZ} \left(S_i^YS_{i+1}^Z + S_i^ZS_{i+1}^Y\right)\Bigr] \Big\}
        \label{eq:results:mag_ham1}
    \end{aligned}
\end{align}
with odd numbers of $i$ on {\it bond 1} and even ones for {\it bond 2}. Within this parametrization $J$ represents the overall scale of the couplings. 

As discussed in detail in Refs~\cite{liu2018pseudospin,liu2020kitaevspinliquid,sano2018kitaevheisenberghamiltonian}, there are two main categories of contributions to the magnetic exchange. The first involves all of those processes that can be downfolded into kinetic exchange processes between $d$-orbital Wannier functions, arising at order $t^2$ in the downfolded $d$-only picture. These terms include the main contribution to the bond-dependent anisotropic exchange terms. The second category of contributions are ligand-exchange terms, involving the mixture of the ground state $j_{1/2}$ configurations with excited states having two holes in different $p$-orbitals on the same ligand. When downfolded into the $d$-orbital basis, such contributions appear as effective intersite Hund's couplings in the electronic Hamiltonian. In the low-energy spin model, they produce ferromagnetic intersite couplings with an anisotropy that is essentially bond-independent. This anisotropy originates from the specific spin-orbital composition of the single-ion states, as revealed in the anisotropy of the $g$-tensor. 

\begin{figure}[b]
    \begin{minipage}[b]{\linewidth}
    \includegraphics[width=.95\linewidth]{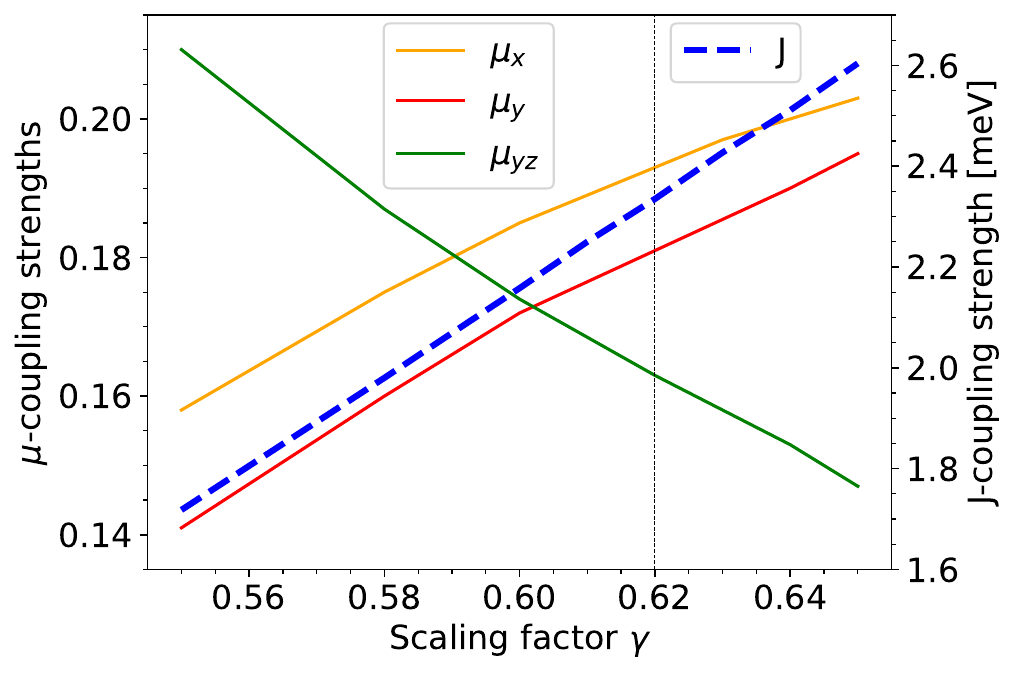}
    \caption{Evolution of main magnetic coupling parameters, defined in  Eq.~\eqref{eq:results:mag_ham1}, with respect to the nearest-neighbor Coulomb screening $\gamma$. The vertical dashed line at $\gamma=0.62$ highlights our choice of the screening parameters for the final set of results.}
    \label{results:fig:results:magnetic_strengths} 
    \end{minipage}
\end{figure}

In our numerical approach, discussed in the previous section, we make an explicit approximation for the downfolded intersite Coulomb terms, the strength of which is controlled by a scaling parameter $\gamma$, such that $\mathcal{H}_{\rm nn-U} \propto \gamma$. This represents the renormalization of the ligand $p$-orbital Coulomb interactions relative to free ion values. Over a range of different $3d$ compounds, we have found empirically that $\gamma$ values in the range 0.6-0.8 typically reproduce well experimental responses (see, for example, \cite{dhakal2024hybrid}). We therefore perform the projED calculation of the low-energy couplings for a range of $\gamma$ values. Fig.~\ref{results:fig:results:magnetic_strengths} displays the development of the magnetic couplings with varying $\gamma$ for the range $\gamma=0.55$ to $0.65$. Here we considered the parametrization given in Eq.~\ref{eq:results:mag_ham1}. The nearest neighbor couplings are always ferromagnetic, with a dominant Ising anisotropy for the entire range of reasonable $\gamma$ values. $\mu_{XY}$ and $\mu_{ZY}$ have matching signs that alternate from bond to bond while $\mu_{XZ}$ is overall positive. Following Fig.~\ref{results:fig:results:magnetic_strengths} and Table~\ref{appendixtab:results:magnetic_strengths}, between $\gamma = 0.60$ and 1, $\mu_X$ and $\mu_Y$ range between roughly 0.18 and 0.26, which represents a relatively narrow range. $\mu_{YZ}$ varies to a somewhat larger degree, between roughly 0.16 and 0.09.

Thus, we find that when starting from an electronic Hamiltonian able to reproduce the main properties of the $g$-tensor, we are able to reconstruct the ferromagnetic Ising nature of the couplings and provide a relatively narrow microscopically compatible range for the smaller non-Ising terms, even when allowing $\gamma$ to vary.

\begin{table}[t]
    \centering
    \begin{tabular}{|c|c|c|c|}
        \cline{2-4}
        \multicolumn{1}{c|}{} & \multicolumn{1}{c|}{} & \multicolumn{1}{c|}{} & \multicolumn{1}{c|}{} \\[-3pt]
        \multicolumn{1}{c|}{} & Fit$^{\rm INS}$ & Fit$^{\rm THz}$ & $\ \gamma=0.62$, Model$_{\rm corr}^{\rm DFT}\ $ \\[5pt]
        \hline
        & \multicolumn{1}{c|}{} & \multicolumn{1}{c|}{} & \multicolumn{1}{c|}{} \\[-3pt]
        $J_K$    & -0.880 & -0.378 & -0.920 \\[5pt]
        $\bar{K}$   & -0.069 & \phantom{-}0.013 & \phantom{-}0.051 \\[5pt]
        $K$    & -1.020 & -1.159 & -0.499 \\[5pt]
        $\Gamma_1$ & \phantom{-}0.333 & \phantom{-}0.372 & \phantom{-}0.300 \\[5pt]
        $\Gamma_2$ & -0.766 & -0.763 & -0.746 \\[5pt]
        $\Gamma_3$ & \phantom{-}0.665 & \phantom{-}0.749 & \phantom{-}0.820 \\[5pt]
        \hline
    \end{tabular}
    \caption{Comparison of the magnetic exchange parameters reported from fitting to INS data~\cite{coldea2023tuning} (Fit$^{\rm INS}$), THz data~\cite{armitage2021duality} (Fit$^{\rm THz}$) and this work (Model$_{\rm corr}^{\rm DFT}$) for $\gamma=0.62$. The parameters are given in the coordinate system in Fig.~\ref{fig:methods:structure_big_axes}a with the notation in Eq.~\eqref{eq:results:mag_ham1_kitaev}. All couplings are given in [meV].}
    \label{tab:results:magnetic_strenghts_kitaev}
\end{table}

\begin{table}[t]
    \centering
    \begin{tabular}{|c|c|c|c|}
        \cline{2-4}
        \multicolumn{1}{c|}{} & \multicolumn{1}{c|}{} & \multicolumn{1}{c|}{} & \multicolumn{1}{c|}{} \\[-3pt]
        \multicolumn{1}{c|}{} & $\ $Fit$^{\rm INS}\ $ & $\ $Fit$^{\rm THz}\ $ & $\ \gamma=0.62$, Model$_{\rm corr}^{\rm DFT}\ $ \\[5pt]
        \hline
        & \multicolumn{1}{c|}{} & \multicolumn{1}{c|}{} & \multicolumn{1}{c|}{} \\[-3pt]
        $J$        & 2.48(2) & 2.085 & 2.335 \\[5pt]
        $\mu_X$    & 0.251(6) & 0 & 0.193 \\[5pt]
        $\mu_Y$    & 0.251(6) & 0.0935 & 0.181 \\[5pt]
        $\mu_{XY}$ & 0 & 0 & 0.098 \\[5pt]
        $\mu_{YZ}$ & 0.226(3) & 0.306 & 0.163 \\[5pt]
        $ \mu_{XZ} $ & 0 & 0 & 0.050 \\[5pt]
        \hline
    \end{tabular}
    \caption{Comparison of the magnetic exchange parameters reported from fitting
    to INS data~\cite{coldea2023tuning} (Fit$^{\rm INS}$), THz data~\cite{armitage2021duality} (Fit$^{\rm THz}$) and this work
    (Model$_{\rm corr}^{\rm DFT}$) for $\gamma=0.62$. The parameters are given in the coordinate system in Fig.~\ref{fig:methods:structure_big_axes}b with the notation in Eq.~\eqref{eq:results:mag_ham1}. All couplings are given in [meV].}
    \label{tab:results:magnetic_strenghts}
\end{table}
In Table~\ref{tab:results:magnetic_strenghts_kitaev} and ~\ref{tab:results:magnetic_strenghts} we display the calculated magnetic exchange parameters in the coordinate systems presented in Fig.~\ref{fig:methods:structure_big_axes} {\bf a} and Fig.~\ref{fig:methods:structure_big_axes} {\bf b} for a screening parameter $\gamma = 0.62$. This choice allows for a good reconstruction of the overall bandwidth of the measured excitation spectrum as shown in the next section. Table~\ref{tab:results:magnetic_strenghts_kitaev} and ~\ref{tab:results:magnetic_strenghts} also include the exchange interactions obtained in Refs.~\cite{coldea2023tuning} and ~\cite{armitage2021duality} from fitting to experimental data. For this purpose we transformed the notation in~\cite{coldea2023tuning,armitage2021duality} into the coordinate systems used in this work.

Within the principal axis system (\textbf{XYZ}, see Table \ref{tab:results:magnetic_strenghts}), following the fitted results, $J$ is expected to be at $\approx$2-2.5meV. While the $Z$-component exhibits the full influence of $J$, the $X$ and $Y$-components are diminished by a significant amount. While Ref.~\cite{coldea2023tuning} expects the $S_XS_X$ and $S_YS_Y$ components to be similar at $\approx$0.25$J$, Ref.~\cite{armitage2021duality} proposes a value of $\approx$0.19$J$ for the $S_YS_Y$ component and no contribution at all from the $S_XS_X$ component. Our data shows a trend similar to ``Fit$^{\rm INS}$'', though with slightly smaller values of $\mu_X$ and $\mu_Y$.

While our calculations predict small magnitudes for both $\mu_{XY}$ and $\mu_{XZ}$, both fitting procedures set them to zero. Possibly, within the fitted models the effects of finite $\mu_{XY}$ and $\mu_{XZ}$ have been reabsorbed into an increased effective $\mu_{YZ}$. Compared to our results, $\mu_{YZ}$ is indeed slightly larger in both experimental fittings. In conclusion, we observe that our calculated parameters have the strongest resemblance to those obtained in Ref.~\cite{coldea2023tuning}.

Additional data was collected in \nameref{appendix:detailed_results} Table~\ref{appendixtab:results:magnetic_strengths} to visualize the influence of the parameters in the Hubbard model on the couplings. Couplings resulting from Model$^{\rm DFT}$ and Model$_{\rm corr}^{\rm DFT}$ as well as for $\gamma=0$, 0.6 and 1 are presented. While $\mu_{XY}$ is very small in general, the orders of magnitude of $\mu_{XZ}$ and $\mu_{YZ}$ vary strongly with the underlying model parameters. $\mu_{XZ}$ changes depending on whether or not the adjustment to the crystal field is applied. Adding screening to the nearest-neighbor Coulomb interaction raises $\mu_{YZ}$ to values at the same level as the diagonal elements $\mu_X$ and $\mu_Y$.

In Table \ref{appendixtab:results:magnetic_strengths_plus} in \nameref{appendix:detailed_results} a collection of magnetic couplings obtained through a selective choice of various contributions in the electronic Hamiltonian is presented. This allows for a deeper understanding of the influence of individual terms and the origin of different exchange terms. For example, the case of an ideal octahedral crystal field has been investigated. In that case, all anisotropies originating from the crystal field can be ruled out and other potential sources of anisotropic couplings can be uncovered. Similarly, as has been done before, by setting $\gamma=0$ the ligand-mediated Hund's exchange can be turned off while the exact influence of the kinetic exchange can be made visible by setting intersite hoppings to zero.

Based on the results the following conclusions can be made: \\
(i) In the absence of crystal field distortions the ligand-mediated Hund's exchange exhibits ferromagnetic and nearly isotropic behavior, while the kinetic exchange is antiferromagnetic and displays a significant anisotropy. However, large direct hoppings between $d$-orbitals on adjacent sites prevent the anisotropies from taking the Kitaev form (which arise primarily from ligand-mediated hoppings). \\
(ii) In the presence of the crystal field distortion, both, the kinetic and ligand-mediated Hund’s couplings are rendered anisotropic. Overall, the distorted crystal field now predominantly contributes to the anisotropy of the couplings by displaying dominant XXZ-type anisotropic behavior of similar magnitude in both cases. \\
(iii) The final couplings emerge as a compromise of these contributions. The larger magnitude of the ferromagnetic ligand-mediated exchange term with respect to the antiferromagnetic hopping term leads to the overall ferromagnetic Ising character of the system. \\
(iv) The smaller $\mu_{XY}$, $\mu_{YZ}$ and $\mu_{XZ}$ terms mostly originate from the kinetic exchange.

\vspace{0.2cm}

\begin{figure}
    \begin{minipage}[b]{\linewidth}
    \includegraphics[width=1\linewidth]{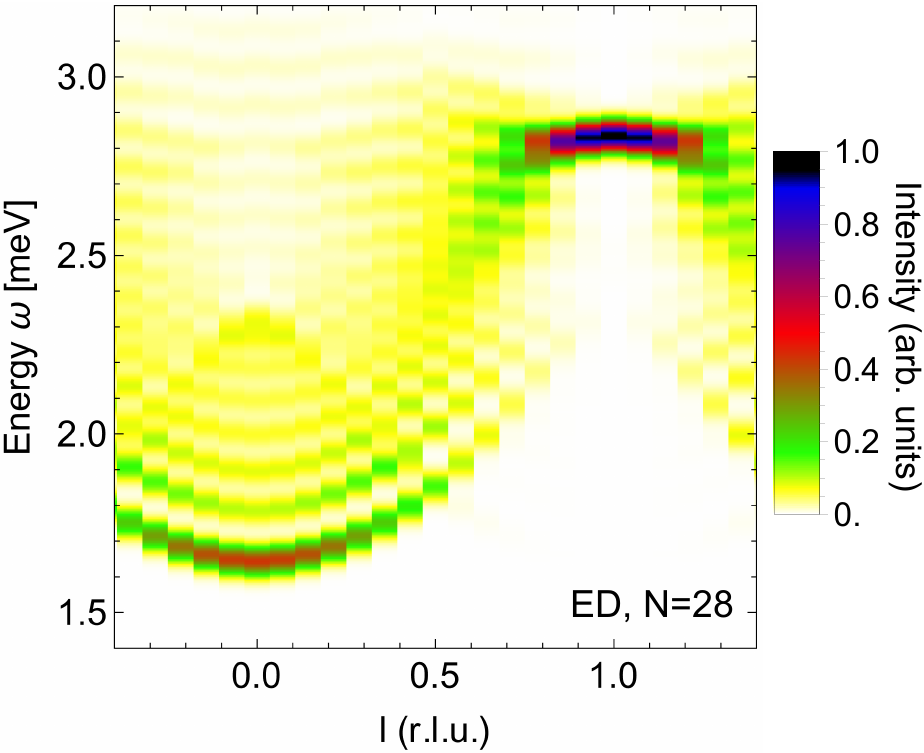}
    \caption{\label{fig:INS}
    $\mathcal S^{XX}(\mathbf q,\omega)$ for the exchange parameters given in \cref{tab:results:magnetic_strenghts} ($\gamma=0.62$, Model$_\text{corr}^\text{DFT}$) at zero magnetic field computed via exact diagonalization on a periodic chain of $N=28$ sites. The color function is oriented at the one of the experimental INS data in Ref.\cite{coldea2023excitationsofquantumisingchain}.}
   \end{minipage}
\end{figure}

\begin{figure}
   \begin{minipage}[b]{\linewidth}
    \centering
    \includegraphics[width=0.995\linewidth]{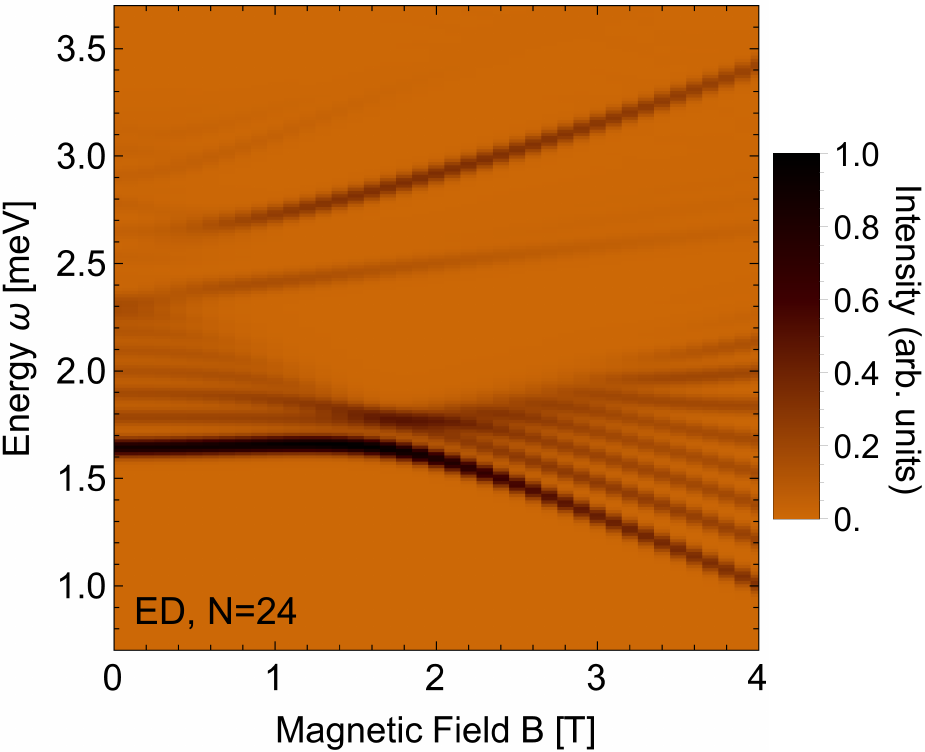}
    \caption{\label{fig:ESR}
    $\mathcal S^{XX}(\mathbf q=0,\omega)$ as a function of transverse magnetic field $B$ ($\mathbf B\parallel b$) for the exchange parameters given in \cref{tab:results:magnetic_strenghts} (``$\gamma=0.62$, Model$_{\rm corr}^{\rm DFT}$'') and $g_Y=3.13$ from \cref{tab:results:gtensor}. Computed via exact diagonalization on a periodic chain of $N=24$ sites. The color function is oriented at the one of the experimental THz data in Ref.~\cite{armitage2021duality}.}
    \end{minipage}
\end{figure}


\section{DISCUSSION}

\vspace{0.2cm}

To benchmark our \textit{ab-initio} procedure and the derived
spin exchange model, we perform exact diagonalization calculations of the $j_{1/2}$ model in Eq.~\eqref{eq:results:mag_ham1} and compare to INS~\cite{coldea2023tuning} and field-dependent THz spectroscopy data~\cite{armitage2021duality}. We therefore analyze the zero-temperature dynamical spin-structure factor 
\begin{align}
    \mathcal S^{XX}(\mathbf q,\omega) = \sum_{n} \left|\braket{0|S^X_\mathbf{q}|n}\right|^2 \delta(E_n-E_0-\omega), 
\end{align}
which we compute via the method of Ref.~\cite{dagotto1994correlated,winter2018probing} using 150 Lanczos vectors for the spectral decomposition. Results for INS and THz data are presented in \cref{fig:INS,fig:ESR}, respectively, both with a Gaussian broadening of $\sigma=0.028\,$meV. \\
The weak inter-chain couplings leading to antiferromagnetic order between the chains are incorporated on a mean-field level through the addition of a simple term $\mathcal H_\text{MF} = -\sum_i h_\text{MF} S_i^Z $ to the Hamiltonian, for which we choose $h_\text{MF}=0.04\,$meV, oriented at the strength of mean-fields in previous studies \cite{coldea2023excitationsofquantumisingchain,armitage2021duality}. 
Within ED, we find in the $\gamma=0.62$ model the ordered moment $\braket{\mathbf g\cdot\mathbf S}$ to be within $1^\circ$ off the {\bf Z}-direction at $B=0$, consistent with experiment \cite{heid1995magphasediag,kohn1999threedimensional,wanklyn1979magneticstructures,SetsuoMitsuda1994magordering}. 

The dynamical response computed within ED of our first-principles derived exchange model (Model$_{\rm corr}^{\rm DFT}$ with $\gamma$=0.62) is in good agreement with the available experimental data \cite{coldea2023excitationsofquantumisingchain,armitage2021duality}. 
In particular, the zero-field INS response [\cref{fig:INS}] reproduces both the sharp high-energy mode near $l=1$ and the broad scattering continuum away from $l=1$ \cite{coldea2023excitationsofquantumisingchain}. The latter is induced by the off-diagonal non-Ising exchanges and can be understood as originating from a finite dispersion being lent to the domain-wall excitations \cite{fava2020glidesymmetrybreaking,armitage2021duality}. 
Near $l=0$, the series of discrete bound states is reproduced \cite{coldea2023excitationsofquantumisingchain}, which are induced by the longitudinal Weiss field from interchain coupling, that acts effectively as a confining potential for the domain wall excitations \cite{coldea2010quantumcriticalityinanisingchain,morris2014hierarchyofboundstates}. The field-dependence seen in the THz spectrum [\cref{fig:ESR}] is also qualitatively consistent with experiment \cite{armitage2021duality}, featuring  for the low-energy excitations a minimum of the excitation bandwidth near $B=2\,$T, for the medium-energy excitation a weaker $B$-dependence, and for the high-energy excitation a larger positive slope with field. Nevertheless, the relative intensities of the observed modes are not in perfect agreement with experiment \cite{armitage2021duality}. Below, we discuss such quantitative deviations from the experiments in detail and provide arguments on their origin.  

The most prominent difference is the deviation in the overall energy scale which is slightly smaller in our \textit{ab-initio} derived model compared to experiment. The peak visible in the INS spectrum displayed in Fig.~\ref{fig:INS} lies at approximately 2.85meV while the data in \cite{coldea2023tuning} suggest an experimentally found energy of about 3.1meV for the peak. The ESR spectrum is shifted equivalently. This energy scale is mainly incorporated in the size of the coupling $J$, which has a strong dependence on the screening of the nearest neighbor Coulomb terms. Smaller screening would increase $J$ though this in turn would lead to the worsening of other aspects of the data. Other factors impacting $J$ are also the magnitudes of the $U$ and $J_H$ Coulomb interactions on a single site. For our computation we used the parameters calculated in \cite{scheffler2010firstprinciplesmodeling} though also other values can be found in older literature, e.g.\ in \cite{turner1991electronicLiCoO2}. Fig.~\ref{fig:INS} also shows that the width of our energy distribution is smaller than in \cite{coldea2023tuning}. We found the size of the off-diagonal exchange couplings to be crucial for the width of the energy scale. Though larger off-diagonal contributions would have broadened the width of $\omega$, this adjustment wasn't possible within our framework.

Another noteworthy difference between our data and experiment is visible in the ESR spectrum (see Fig.~\ref{fig:ESR}). Here the lowest lying energy mode has the strongest intensity which isn't the case in the experimental findings in \cite{armitage2021duality}. The intensity distribution could be adapted to the experimental ESR spectrum through increasing off-site Coulomb screening though this would lower the energy scale further.

Summarizing, the ESR and THz spectra extracted from our exact diagonalization approach of the first-principles modelled Hamiltonian, are able to reproduce experimental findings on a qualitative and also partially on a quantitative level. We were also able to determine the microscopic origins to particular aspects of the spectra like nearest-neighbor Coulomb screening and on-site Coulomb interaction strengths. While we obtained an overall good qualitative agreement, note that we have not taken into account further-neighbor exchanges, which are likely relevant for a further improved quantitative agreement \cite{coldea2023excitationsofquantumisingchain}.

\section{CONCLUSIONS}

In this work, we have shown that the different refined models for the Ising spin-chain compound CoNb$_2$O$_6$ derived recently from experimental fitting \cite{armitage2021duality,coldea2023excitationsofquantumisingchain,coldea2023tuning,churchill2024transforming} offer a mutually compatible view of the material that is consistent with microscopic considerations. Specifically, in this material, the local Co spins are strongly perturbed $j_{1/2}$ moments of the high-spin $d^7$ configuration. The largest and most crucial contribution to the magnetic exchange is the ferromagnetic Goodenough-Kanamori (ligand-mediated) exchange involving the $e_g$ electrons. This contribution acquires a bond-independent XXZ (Ising) anisotropy as a consequence of the strongly distorted CoO$_6$ octahedra, which also results in anisotropic $g$-tensors. The smaller bond-dependent couplings, the importance of which have been highlighted by the recent neutron and THz studies, arise as residual effects of $d-d$ kinetic exchange processes involving the $t_{2g}$ electrons. However, these residual terms do not follow precisely a Kitaev form because of the effects of the crystal field distortion, and the specific magnitudes of the $d-d$ hoppings are more conducive to off-diagonal $\Gamma$ couplings. These findings are all compatible with recent assertions about couplings between high-spin $d^7$ compounds in \cite{winter2022magneticcouplings}, and offer significant insight into related Co$^{2+}$ compounds currently under study in the context of bond-dependent magnetism.

In order to derive these conclusions, we have followed an {\it ab-initio} based approach that allows for a balanced treatment of the relevant contributions to the magnetic couplings, including the ligand-mediated ferromagnetic exchange, within a $d$-orbital picture. By adjusting the crystal field to match the experimental $g-$tensor, and varying the strength of the intersite Coulomb exchange, we have identified a relatively narrow range of magnetic Hamiltonians that can be considered microscopically compatible, and shown that they reproduce the experimentally observed rich dynamical response of the material. Specifically, for the bond-independent XXZ couplings, the two perpendicular (X, Y) contributions are smaller than the Ising (Z) coupling by a factor of roughly $\mu_X\approx\mu_Y\approx 0.2$. The largest bond-dependent coupling is of similar magnitude, and takes the form of off-diagonal $\mu_{YZ}$. As pointed out in \cite{coldea2023excitationsofquantumisingchain}, additional off-diagonal $\mu_{XY}$ and $\mu_{XZ}$ are also symmetry allowed, but were set to zero in both Ref.~\cite{coldea2023tuning} and Ref.~\cite{armitage2021duality} for simplicity. We find these couplings to be finite, but that they are the smallest terms in the Hamiltonian, partially justifying their omission in these previous studies. Overall, our derived couplings most closely resemble those of Ref.~\cite{coldea2023tuning}, the main distinction being that Ref.~\cite{coldea2023tuning} employs $\mu_X = \mu_Y$, while Ref.~\cite{armitage2021duality} employs $\mu_X = 0$.
Thus, while CoNb$_2$O$_6$ remains a very complex system to investigate within a purely \textit{ab-initio} approach, the utilization of experimental input as guidance (such as the $g$-tensors and overall magnetic excitation bandwidth), allows for the elucidation of a microscopically consistent picture of the varied and competing exchange contributions. After many years of intensive research, the description of this paradigmatic material in quantum magnetism is finally coalescing, and continuing to hold new insights.

\vspace{0.2cm}
\section{ACKNOWLEDGEMENTS}

\vspace{0.2cm}

We would like to thank R.~Coldea, P.~Armitage, H.-Y.~Kee and A.~L.~Chernyshev for valuable discussions.
We acknowledge support by the Deutsche Forschungsgemeinschaft (DFG, German Research Foundation) for funding through project TRR 288 — 422213477 (project A05) and through QUAST-FOR5249 - 449872909 (project TP4). 

\clearpage

\appendix

\section{Appendix A}\label{appendix:groundstateprojection}
\subsection{Ground state multiplet projection}

Within the projED approach the Hamiltonian from Eq.~\eqref{eq:methods:general_hamiltonian} is diagonalized numerically for one or two site clusters, retaining the lowest doublet state(s). The resulting low-energy Hamiltonian in the basis of these doublets is then projected onto an idealized low-energy space of pure $j_{1/2}$ doublets. The main contributions to these states are within the ($t_{2g}$)$^5$($e_g$)$^2$ subspace with seven electrons per site and $J_{eff}=1/2$  \cite{winter2022magneticcouplings}. The aim is to project the initially in orbital-basis given states onto the $J_{eff}=1/2$ single-ion doublets. This is done in two steps. First, for the material at hand the orbital basis can rewritten in terms of $\ket{m_L,m_S}$ states. Those compose the $j_{1/2}$-basis in the following way \cite{lines1963magneticproperties}:
\begin{subequations}
    \begin{align}
        & \ket{\frac{1}{2},+\frac{1}{2}} = c_1\ket{-1,\frac{3}{2}} + c_2\ket{0,\frac{1}{2}} + c_3\ket{1,-\frac{1}{2}} \label{eq:methods:groundstate_doublet1} \\
        & \ket{\frac{1}{2},-\frac{1}{2}} = c_1\ket{1,-\frac{3}{2}} + c_2\ket{0,-\frac{1}{2}} + c_3\ket{-1,\frac{1}{2}} \label{eq:methods:groundstate_doublet2}
    \end{align}
\end{subequations}
where the prefactors $c_1$, $c_2$ and $c_3$ can be chosen as $c_1=1/\sqrt{2}$, $c_2=1/\sqrt{3}$ and $c_3=1/\sqrt{6}$. This actually refers to to the case of zero trigonal splitting \cite{lines1963magneticproperties,winter2022magneticcouplings}, though variation in this coefficients would only result in Hamiltonians that are related through unitary basis transformations. Since the calculation of both the magnetic couplings and $g$-tensors employ the same projection scheme, such unitary transformations do no result in any difference in the local spectrum of the model. These are a composition of orbital states in the $t_{2g}$-$e_g$ basis and can be expressed via:
\begin{subequations}
    \begin{align}
        \ket{-1,\frac{3}{2}} = \ket{e_{a,\uparrow}e_{b,\uparrow}t_{+,\uparrow}t_{0,\uparrow}t_{0,\downarrow}t_{-,\uparrow}t_{-,\downarrow}}
    \end{align}
    \begin{align}
        \begin{aligned}
            \ket{0,\frac{1}{2}} = & \frac{1}{\sqrt{3}} \ket{e_{a,\uparrow}e_{b,\uparrow}t_{+,\uparrow}t_{+,\downarrow}t_{0,\downarrow}t_{-,\uparrow}t_{-,\downarrow}} \\
            & + \frac{1}{\sqrt{3}} \ket{e_{a,\uparrow}e_{b,\downarrow}t_{+,\uparrow}t_{+,\downarrow}t_{0,\uparrow}t_{-,\uparrow}t_{-,\downarrow}} \\
            & + \frac{1}{\sqrt{3}} \ket{e_{a,\downarrow}e_{b,\uparrow}t_{+,\uparrow}t_{+,\downarrow}t_{0,\uparrow}t_{-,\uparrow}t_{-,\downarrow}}
        \end{aligned}
    \end{align}
    \begin{align}
        \begin{aligned}
            \ket{1,-\frac{1}{2}} = & \frac{1}{\sqrt{3}} \ket{e_{a,\uparrow}e_{b,\downarrow}t_{+,\uparrow}t_{+,\downarrow}t_{0,\uparrow}t_{0,\downarrow}t_{-,\downarrow}} \\
            & + \frac{1}{\sqrt{3}} \ket{e_{a,\downarrow}e_{b,\uparrow}t_{+,\uparrow}t_{+,\downarrow}t_{0,\uparrow}t_{0,\downarrow}t_{-,\downarrow}} \\
            & + \frac{1}{\sqrt{3}} \ket{e_{a,\downarrow}e_{b,\downarrow}t_{+,\uparrow}t_{+,\downarrow}t_{0,\uparrow}t_{0,\downarrow}t_{-,\uparrow}}
        \end{aligned}
    \end{align}
\end{subequations}
Since within FPLO we worked within the basis of $d_{xy}$, $d_{yz}$, $d_{xz}$, $d_{z^2}$ and $d_{x^2-y^2}$ orbitals, a transformation between this states and the  $t_{2g}$-$e_g$ states has to be taken into account:
\begin{subequations}
    \begin{align}
        \ket{e_{a,\sigma}} = \ket{d_{z^2,\sigma}}
    \end{align}
    \begin{align}
        \ket{e_{b,\sigma}} = \ket{d_{x^2-y^2,\sigma}}
    \end{align}
    \begin{align}
        \ket{t_{+,\sigma}} = -\frac{1}{\sqrt{2}} \left( \ket{d_{yz,\sigma}}+i\ket{d_{xz,\sigma}} \right)
    \end{align}
    \begin{align}
        \ket{t_{0,\sigma}} = \ket{d_{xy,\sigma}}
    \end{align}
    \begin{align}
        \ket{t_{-,\sigma}} = \frac{1}{\sqrt{2}} \left( \ket{d_{yz,\sigma}}-i\ket{d_{xz,\sigma}} \right)
    \end{align}
\end{subequations}
The time-reversed states can be obtained analogously.

\vspace{0.2cm}

\section{Appendix B}\label{appendix:hopping_NNCoulomb_parameters}
\subsection{Hopping and nearest-neighbor Coulomb parameters}
\vspace{0.2cm}

The hoppings in meV on {\it bond 2} (the bond perpendicular to the $\mathbf z$-axis) between nearest-neighbors determined by FPLO are given, in terms of $d_{xy},d_{yz},d_{xz},d_{z^2}$ and $d_{x^2-y^2}$ orbitals, as
\begin{align}
    \begin{pmatrix}
        -69.95 & -2.34 & 21.88 & 88.93  & -5.33  \\
        -2.34  & 24.81 & 40.78 & -5.52  & 16.56  \\
        21.88  & 40.78 & 11.32 & 56.00  & -5.56  \\
        88.93  & -5.52 & 56.00 & -80.96 & 2.58   \\
        -5.33  & 16.56 & -5.56 & 2.58   & -55.66 \\
    \end{pmatrix}
\end{align}
The matrices accounting for nearest-neighbor Coulomb interaction on {\it bond 2} are (in meV):
\begin{subequations}
\label{eq:UUJmatrices}
\begin{align}
    \begin{aligned}
        \tilde U_1 = (U_0-2J_H) & A_{ab} + 2J_HB_{ab} + U_0C_{ab} \\ = &  \ \begin{pmatrix}
        0.38 & 0.23 & 0.31 & 0.75 & 2.47 \\
        0.23 & 0    & 0.60 & 0.23 & 0.79 \\
        0.31 & 0.69 & 0    & 0.21 & 0.58 \\
        0.75 & 0.23 & 0.21 & 0.32 & 1.04 \\
        2.47 & 0.79 & 0.58 & 1.04 & 3.33 \\
        \end{pmatrix}
    \end{aligned}
\end{align}
\begin{align}
    \begin{aligned}
        \tilde U_2 = \frac{U_0-3J_H}{2} & \left(A_{ab} - B_{ab}\right) \\ = &  \ \begin{pmatrix}
        0.04 & 0.02 & 0.02 & 0.01 & 0.01 \\
        0.02 & 0    & 0.01 & 0.03 & 0.10 \\
        0.02 & 0.01 & 0    & 0.03 & 0.07 \\
        0.01 & 0.03 & 0.03 & 0.04 & 0.13 \\
        0.01 & 0.10 & 0.07 & 0.13 & 0.39 \\
        \end{pmatrix}
    \end{aligned}
\end{align}
\begin{align}
    \begin{aligned}
        \tilde J = J_HA_{ab} + & (U_0-J_H) B_{ab} + U_0C_{ab} \\ = & \ \begin{pmatrix}
        0.31 & 0.18 & 0.26 & 0.74  & 2.45 \\
        0.18 & 0    & 0.58 & 0.18  & 0.60 \\
        0.26 & 0.58 & 0    & 0.16  & 0.44 \\
        0.74 & 0.18 & 0.16 & 0.24  & 0.79 \\
        2.45 & 0.60 & 0.44 & 0.79  & 2.54 \\
        \end{pmatrix}
    \end{aligned}
\end{align}
\end{subequations}

\vspace{0.2cm}

\section{Appendix C}\label{appendix:detailed_results}
\subsection{Detailed results on $g$-tensor and Spin Hamiltonian}
\vspace{0.2cm}

$g$-tensors in the $xyz$-coordinate system visible in Fig.~\ref{fig:methods:structure_big_axes}a obtained from Model$^\text{DFT}$:
\begin{align}
    \begin{aligned}
        g = \begin{pmatrix}
            5.212 & -0.553 & 1.666 \\
            -0.553 & 2.324 & -0.551 \\
            1.670 & -0.551 & 5.208
        \end{pmatrix}
    \end{aligned}
    \label{appendixeq:gtensor1}
\end{align}
and Model$_\text{corr}^\text{DFT}$:
\begin{align}
    \begin{aligned}
        g = \begin{pmatrix}
            4.697 & -1.033 & 1.564 \\
            -1.033 & 3.629 & -1.033 \\
            1.566 & -1.033 & 4.684
        \end{pmatrix}
    \end{aligned}
    \label{appendixeq:gtensor2}
\end{align}
The principle values and axes for both models are visible in Table \ref{tab:results:principal_values_axes}. The eigenvalues for Model$_\text{corr}^\text{DFT}$ are compared to fitted data in Table \ref{tab:results:gtensor} in the main text.

The nearest-neighbor Coulomb interaction consists of a spin-independent and a spin-dependent part of which the latter one has the major contribution to the Spin Hamiltonian. Omission of the spin-independent part, as described in the main text, leads to the following magnetic couplings:
\begin{align}
    \begin{aligned}
        J = & 2.332 \\
        \mu_X = & 0.193 \\
        \mu_Y = & 0.183 \\
        \mu_{XY} = & 0.098 \\
        \mu_{YZ} = & 0.158 \\
        \mu_{XZ} = & 0.047 \\
    \end{aligned}
\end{align}
All deviations of the result with full $\mathcal H_{\rm nn-U}$ are of an order of magnitude of $10^{-3}$meV.

\begin{table}[t]
    \centering
    \begin{tabular}{|c|c|c|}
        \cline{2-3}
        \multicolumn{1}{c|}{} & \multicolumn{1}{c|}{} & \multicolumn{1}{c|}{} \\[-3pt]
        \multicolumn{1}{c|}{} & Model$^\text{DFT}$  &  Model$_\text{corr}^\text{DFT}$ \\[5pt]
        \hline
        \multicolumn{1}{|c|}{} & \multicolumn{1}{c|}{} & \multicolumn{1}{c|}{} \\[-3pt]
        $g_X$ & 2.19 & 2.98 \\[5pt]
        $g_Y$ & 3.52 & 3.13 \\[5pt]
        $g_Z$ & 7.01 & 6.90 \\[5pt]
        \hline
        \multicolumn{1}{|c|}{} & \multicolumn{1}{c|}{} & \multicolumn{1}{c|}{} \\[-3pt]
        $\quad \vec v_X\quad$ & $\left( 0.70,0,0.71 \right)$ & $\left( 0.86,-0.01,0.51 \right)$ \\[5pt]
        $\vec v_Y$ & $\left( 0,-1,0 \right)$ & $\left( 0,-1,0 \right)$ \\[5pt]
        $\vec v_Z$ & $\left( 0.71,0,-0.70 \right)$ & $\left( 0.51,0,-0.86 \right)$ \\[5pt]
        \hline
    \end{tabular}
    \caption{$g$-tensor principal values and axes in terms of crystallographic coordinates when considering the crystal field matrix Eq.~\ref{eq:methods:CFS_matrix} (Model$^\text{DFT}$) and the corrected crystal field matrix (Model$_\text{corr}^\text{DFT}$). The axes of Model$^\text{DFT}$ are close to \cref{eq:coordinates_XYZ} for $\phi\approx 45^\circ$, while those of Model$_\text{corr}^\text{DFT}$ are close to $\phi\approx 30^\circ$.}
    \label{tab:results:principal_values_axes}
\end{table}

Detailed Spin Hamiltonian results for Model$^{\rm DFT}$ and Model$_{\rm corr}^{\rm DFT}$ as well as several values of $\gamma$ are visible in Table~\ref{appendixtab:results:magnetic_strengths}. There, the following scenarios were investigated:
\begin{itemize}
    \item[1.] Using CFS parameters directly from FPLO (Model$^{\rm DFT}$) and omitting the influence of the nearest-neighbor Coulomb interaction completely ($\gamma=0$).
    \item[2.] Working with Model$^{\rm DFT}$ and the nearest-neighbor Coulomb interaction defined in in Eq.~\eqref{eq:methods:nnHamiltonian_short}, therefore effectively using a prefactor $\gamma=1$.
    \item[3.] Using adjusted CFS parameters with reduced xy/yz-xz orbital splitting (Model$_{\rm corr}^{\rm DFT}$) and again the nearest-neighbor Coulomb interaction from Eq.~\eqref{eq:methods:nnHamiltonian_short} ($\gamma=1$).
    \item[4.] Using Model$^{\rm DFT}$ and the nearest-neighbor Coulomb interaction from Eq.~\eqref{eq:methods:nnHamiltonian_short}, this time with a prefactor $\gamma=0.62$ to $\mathcal H_{\rm nn-U}$ in order to account for screening effects.
    \item[5.] Working with Model$_{\rm corr}^{\rm DFT}$ while the nearest-neighbor Coulomb interaction $\mathcal H_{\rm nn-U}$ shall again be scaled with a prefactor $\gamma=0.62$.
\end{itemize}

\begin{table*}[t]
    \centering
    \begin{tabular}{|c|c|c|c|c|c|c|c|}
        \cline{2-8}
        \multicolumn{1}{c|}{} & \multirow{2}{*}{Fit$^{\rm INS}$} & \multirow{2}{*}{Fit$^{\rm THz}$} & $\gamma=0$ & \multicolumn{2}{c|}{$\gamma=1$} & \multicolumn{2}{c|}{$\gamma=0.62$} \\
        \multicolumn{1}{c|}{} &  &  & Model$^{\rm DFT}$ & \multicolumn{1}{c}{Model$^{\rm DFT}$} & \multicolumn{1}{c|}{Model$_{\rm corr}^{\rm DFT}$} & \multicolumn{1}{c}{Model$^{\rm DFT}$} & \multicolumn{1}{c|}{Model$_{\rm corr}^{\rm DFT}$} \\
        \hline
        & \multicolumn{1}{c|}{} & \multicolumn{1}{c|}{} & \multicolumn{1}{c|}{} & \multicolumn{1}{c|}{} & \multicolumn{1}{c|}{} & \multicolumn{1}{c|}{} & \multicolumn{1}{c|}{} \\[-3pt]
        $J$        & 2.48(2) & 2.085 & -3.309 & 4.731 & 5.655 & 1.713 & 2.335 \\[5pt]
        $\mu_X$    & 0.251(6) & 0 & 0.267 & 0.270 & 0.262 & 0.204 & 0.193 \\[5pt]
        $\mu_Y$    & 0.251(6) & 0.0935 & 0.380 & 0.323 & 0.260 & 0.251 & 0.181 \\[5pt]
        $\mu_{XY}$ & 0 & 0 & -0.034 & -0.001 & 0.034 & 0.025 & 0.098 \\[5pt]
        $\mu_{YZ}$ & 0.226(3) & 0.306 & -0.054 & 0.088 & 0.086 & 0.182 & 0.163 \\[5pt]
        $\mu_{XZ}$ & 0 & 0 & -0.247 & -0.230 & 0.017 & -0.235 & 0.050 \\[5pt]
        \hline
    \end{tabular}
    \caption{Comparison of the magnetic exchange parameters reported from fitting to INS data \cite{coldea2023tuning} (Fit$^{\rm INS}$), THz data \cite{armitage2021duality} (Fit$^{\rm THz}$) and this work. We present five different sets of results obtained within our approach. Model$^{\rm DFT}$ refers to calculations with the CF matrix from Eq.~\ref{eq:methods:CFS_matrix} and Model$_{\rm corr}^{\rm DFT}$ to calculations with the corrected CF. The variables are named following the notation in Eqs.~\eqref{eq:Hspin2} and ~\eqref{eq:results:mag_ham1} and are given in [meV].}
    \label{appendixtab:results:magnetic_strengths}
\end{table*}

Additional calculation for different microscopic Hamiltonians were performed in order to be able to distinguish their isolated contributions to the full Hamiltonian of the final model (see Table \ref{appendixtab:results:magnetic_strengths_plus}). Here Model$_{\rm CF_0}^{\rm DFT}$ considers only a minimal modelling of the crystal field which includes solely a homogeneous $e_g$-$t_{2g}$ splitting without all other anisotropies. Model$_{\rm corr}^{\rm DFT}$ is defined as usual. For this two models the following cases were investigated:
\begin{itemize}
    \item[1.] $\gamma=0$: The full microscopic Hamiltonian when omitting the intersite Coulomb interaction.
    \item[2.] $\gamma=0.62$, $\mathcal H_{\rm hop}=0$: The full microscopic Hamiltonian when omitting the contribution of hoppings. The intersite Coulomb interaction is introduced with a screening factor of 0.62.
    \item[3.] $\gamma=0.62$: The full microscopic Hamiltonian with the intersite Coulomb interaction contributing with a screening factor of 0.62.
\end{itemize}

\begin{table*}[t]
    \centering
    \begin{tabular}{|c|c|c|c|c|c|c|}
        \cline{2-7}
        \multicolumn{1}{c|}{} & \multicolumn{2}{c|}{$\gamma=0$} & \multicolumn{2}{c|}{$\gamma=0.62$, $\mathcal H_{\rm hop}=0$} & \multicolumn{2}{c|}{$\gamma=0.62$} \\
        \multicolumn{1}{c|}{} & \multicolumn{1}{c}{Model$_{\rm CF_0}^{\rm DFT}$} & \multicolumn{1}{c|}{Model$_{\rm corr}^{\rm DFT}$} & \multicolumn{1}{c}{Model$_{\rm CF_0}^{\rm DFT}$} & \multicolumn{1}{c|}{Model$_{\rm corr}^{\rm DFT}$} & \multicolumn{1}{c}{Model$_{\rm CF_0}^{\rm DFT}$} & \multicolumn{1}{c|}{Model$_{\rm corr}^{\rm DFT}$} \\
        \hline
        & \multicolumn{1}{c|}{} & \multicolumn{1}{c|}{} & \multicolumn{1}{c|}{} & \multicolumn{1}{c|}{} & \multicolumn{1}{c|}{} & \multicolumn{1}{c|}{} \\[-3pt]
        $J$        & -1.829 & -3.197 & 2.874  & 5.411  & 0.973 & 2.335 \\[5pt]
        $\mu_X$    & 1.163  & 0.320  & 1.045  & 0.299  & 0.835 & 0.193 \\[5pt]
        $\mu_Y$    & 1.089  & 0.341  & 0.974  & 0.304  & 0.757 & 0.181 \\[5pt]
        $\mu_{XY}$ & -0.438 & -0.093 & -0.022 & -0.005 & 0.738 & 0.098 \\[5pt]
        $\mu_{YZ}$ & -0.068 & -0.072 & 0.043  & 0.022  & 0.264 & 0.163 \\[5pt]
        $\mu_{XZ}$ & -0.122 & -0.051 & -0.010 & -0.010 & 0.200 & 0.050 \\[5pt]
        \hline
        & \multicolumn{1}{c|}{} & \multicolumn{1}{c|}{} & \multicolumn{1}{c|}{} & \multicolumn{1}{c|}{} & \multicolumn{1}{c|}{} & \multicolumn{1}{c|}{} \\[-3pt]
        $J_K$      & 1.800  & 2.100  & -2.722 & -3.101 & -0.833 & -0.920 \\[5pt]
        $\bar{K}$  & 0.105  & -0.819 & -0.237 & 0.861  & -0.154 & 0.051  \\[5pt]
        $K$        & 0.443  & -0.17  & -0.276 & -0.232 & 0.134  & -0.499 \\[5pt]
        $\Gamma_1$ & -0.368 & -0.767 & -0.044 & 0.978  & -0.405 & 0.300  \\[5pt]
        $\Gamma_2$ & 0.030  & 0.925  & -0.060 & -1.570 & -0.030 & -0.746 \\[5pt]
        $\Gamma_3$ & 0.734  & -0.251 & -0.051 & 1.013  & 0.670  & 0.820  \\[5pt]
        \hline
    \end{tabular}
    \caption{Comparison of magnetic couplings under consideration of different contributions to the microscopic Hamiltonian. Model$_{\rm CF_0}^{\rm DFT}$ represents a minimal modelling for the crystal field, including only a homogeneous $e_g$-$t_{2g}$ splitting. Model$_{\rm corr}^{\rm DFT}$ is defined as before.}
    \label{appendixtab:results:magnetic_strengths_plus}
\end{table*}

\clearpage
%


\begin{thebibliography}{64}%
	\makeatletter
	\providecommand \@ifxundefined [1]{%
		\@ifx{#1\undefined}
	}%
	\providecommand \@ifnum [1]{%
		\ifnum #1\expandafter \@firstoftwo
		\else \expandafter \@secondoftwo
		\fi
	}%
	\providecommand \@ifx [1]{%
		\ifx #1\expandafter \@firstoftwo
		\else \expandafter \@secondoftwo
		\fi
	}%
	\providecommand \natexlab [1]{#1}%
	\providecommand \enquote  [1]{``#1''}%
	\providecommand \bibnamefont  [1]{#1}%
	\providecommand \bibfnamefont [1]{#1}%
	\providecommand \citenamefont [1]{#1}%
	\providecommand \href@noop [0]{\@secondoftwo}%
	\providecommand \href [0]{\begingroup \@sanitize@url \@href}%
	\providecommand \@href[1]{\@@startlink{#1}\@@href}%
	\providecommand \@@href[1]{\endgroup#1\@@endlink}%
	\providecommand \@sanitize@url [0]{\catcode `\\12\catcode `\$12\catcode
		`\&12\catcode `\#12\catcode `\^12\catcode `\_12\catcode `\%12\relax}%
	\providecommand \@@startlink[1]{}%
	\providecommand \@@endlink[0]{}%
	\providecommand \url  [0]{\begingroup\@sanitize@url \@url }%
	\providecommand \@url [1]{\endgroup\@href {#1}{\urlprefix }}%
	\providecommand \urlprefix  [0]{URL }%
	\providecommand \Eprint [0]{\href }%
	\providecommand \doibase [0]{http://dx.doi.org/}%
	\providecommand \selectlanguage [0]{\@gobble}%
	\providecommand \bibinfo  [0]{\@secondoftwo}%
	\providecommand \bibfield  [0]{\@secondoftwo}%
	\providecommand \translation [1]{[#1]}%
	\providecommand \BibitemOpen [0]{}%
	\providecommand \bibitemStop [0]{}%
	\providecommand \bibitemNoStop [0]{.\EOS\space}%
	\providecommand \EOS [0]{\spacefactor3000\relax}%
	\providecommand \BibitemShut  [1]{\csname bibitem#1\endcsname}%
	\let\auto@bib@innerbib\@empty
	\bibitem [{\citenamefont {Maartense}\ \emph {et~al.}(1977)\citenamefont
		{Maartense}, \citenamefont {Yaeger},\ and\ \citenamefont
		{Wanklyn}}]{maartense1977field-inducedmagtrans}%
	\BibitemOpen
	\bibfield  {author} {\bibinfo {author} {\bibfnamefont {I.}~\bibnamefont
			{Maartense}}, \bibinfo {author} {\bibfnamefont {I.}~\bibnamefont {Yaeger}}, \
		and\ \bibinfo {author} {\bibfnamefont {B.}~\bibnamefont {Wanklyn}},\ }\href
	{\doibase https://doi.org/10.1016/0038-1098(77)91485-5} {\bibfield  {journal}
		{\bibinfo  {journal} {Solid State Communications}\ }\textbf {\bibinfo
			{volume} {21}},\ \bibinfo {pages} {93} (\bibinfo {year} {1977})}\BibitemShut
	{NoStop}%
	\bibitem [{\citenamefont {Scharf}\ \emph
		{et~al.}(1979{\natexlab{a}})\citenamefont {Scharf}, \citenamefont {Weitzel},
		\citenamefont {Yaeger}, \citenamefont {Maartense},\ and\ \citenamefont
		{Wanklyn}}]{scharf1979magneticstructures}%
	\BibitemOpen
	\bibfield  {author} {\bibinfo {author} {\bibfnamefont {W.}~\bibnamefont
			{Scharf}}, \bibinfo {author} {\bibfnamefont {H.}~\bibnamefont {Weitzel}},
		\bibinfo {author} {\bibfnamefont {I.}~\bibnamefont {Yaeger}}, \bibinfo
		{author} {\bibfnamefont {I.}~\bibnamefont {Maartense}}, \ and\ \bibinfo
		{author} {\bibfnamefont {B.}~\bibnamefont {Wanklyn}},\ }\href {\doibase
		https://doi.org/10.1016/0304-8853(79)90044-1} {\bibfield  {journal} {\bibinfo
			{journal} {Journal of Magnetism and Magnetic Materials}\ }\textbf {\bibinfo
			{volume} {13}},\ \bibinfo {pages} {121} (\bibinfo {year}
		{1979}{\natexlab{a}})}\BibitemShut {NoStop}%
	\bibitem [{\citenamefont {Hanawa}\ \emph {et~al.}(1994)\citenamefont {Hanawa},
		\citenamefont {Shinkawa}, \citenamefont {Ishikawa}, \citenamefont {Miyatani},
		\citenamefont {Saito},\ and\ \citenamefont
		{Kohn}}]{hanawa1994anisotropicspecificheat}%
	\BibitemOpen
	\bibfield  {author} {\bibinfo {author} {\bibfnamefont {T.}~\bibnamefont
			{Hanawa}}, \bibinfo {author} {\bibfnamefont {K.}~\bibnamefont {Shinkawa}},
		\bibinfo {author} {\bibfnamefont {M.}~\bibnamefont {Ishikawa}}, \bibinfo
		{author} {\bibfnamefont {K.}~\bibnamefont {Miyatani}}, \bibinfo {author}
		{\bibfnamefont {K.}~\bibnamefont {Saito}}, \ and\ \bibinfo {author}
		{\bibfnamefont {K.}~\bibnamefont {Kohn}},\ }\href {\doibase
		10.1143/JPSJ.63.2706} {\bibfield  {journal} {\bibinfo  {journal} {Journal of
				the Physical Society of Japan}\ }\textbf {\bibinfo {volume} {63}},\ \bibinfo
		{pages} {2706} (\bibinfo {year} {1994})}\BibitemShut {NoStop}%
	\bibitem [{\citenamefont {Heid}\ \emph {et~al.}(1995)\citenamefont {Heid},
		\citenamefont {Weitzel}, \citenamefont {Burlet}, \citenamefont {Bonnet},
		\citenamefont {Gonschorek}, \citenamefont {Vogt}, \citenamefont {Norwig},\
		and\ \citenamefont {Fuess}}]{heid1995magphasediag}%
	\BibitemOpen
	\bibfield  {author} {\bibinfo {author} {\bibfnamefont {C.}~\bibnamefont
			{Heid}}, \bibinfo {author} {\bibfnamefont {H.}~\bibnamefont {Weitzel}},
		\bibinfo {author} {\bibfnamefont {P.}~\bibnamefont {Burlet}}, \bibinfo
		{author} {\bibfnamefont {M.}~\bibnamefont {Bonnet}}, \bibinfo {author}
		{\bibfnamefont {W.}~\bibnamefont {Gonschorek}}, \bibinfo {author}
		{\bibfnamefont {T.}~\bibnamefont {Vogt}}, \bibinfo {author} {\bibfnamefont
			{J.}~\bibnamefont {Norwig}}, \ and\ \bibinfo {author} {\bibfnamefont
			{H.}~\bibnamefont {Fuess}},\ }\href {\doibase
		https://doi.org/10.1016/0304-8853(95)00394-0} {\bibfield  {journal} {\bibinfo
			{journal} {Journal of Magnetism and Magnetic Materials}\ }\textbf {\bibinfo
			{volume} {151}},\ \bibinfo {pages} {123} (\bibinfo {year}
		{1995})}\BibitemShut {NoStop}%
	\bibitem [{\citenamefont {Cabrera}\ \emph {et~al.}(2014)\citenamefont
		{Cabrera}, \citenamefont {Thompson}, \citenamefont {Coldea}, \citenamefont
		{Prabhakaran}, \citenamefont {Bewley}, \citenamefont {Guidi}, \citenamefont
		{Rodriguez-Rivera},\ and\ \citenamefont
		{Stock}}]{cabrera2014excitationsinthequantumparamagnetic}%
	\BibitemOpen
	\bibfield  {author} {\bibinfo {author} {\bibfnamefont {I.}~\bibnamefont
			{Cabrera}}, \bibinfo {author} {\bibfnamefont {J.~D.}\ \bibnamefont
			{Thompson}}, \bibinfo {author} {\bibfnamefont {R.}~\bibnamefont {Coldea}},
		\bibinfo {author} {\bibfnamefont {D.}~\bibnamefont {Prabhakaran}}, \bibinfo
		{author} {\bibfnamefont {R.~I.}\ \bibnamefont {Bewley}}, \bibinfo {author}
		{\bibfnamefont {T.}~\bibnamefont {Guidi}}, \bibinfo {author} {\bibfnamefont
			{J.~A.}\ \bibnamefont {Rodriguez-Rivera}}, \ and\ \bibinfo {author}
		{\bibfnamefont {C.}~\bibnamefont {Stock}},\ }\href {\doibase
		10.1103/PhysRevB.90.014418} {\bibfield  {journal} {\bibinfo  {journal} {Phys.
				Rev. B}\ }\textbf {\bibinfo {volume} {90}},\ \bibinfo {pages} {014418}
		(\bibinfo {year} {2014})}\BibitemShut {NoStop}%
	\bibitem [{\citenamefont {Coldea}\ \emph {et~al.}(2010)\citenamefont {Coldea},
		\citenamefont {Tennant}, \citenamefont {Wheeler}, \citenamefont {Wawrzynska},
		\citenamefont {Prabhakaran}, \citenamefont {Telling}, \citenamefont
		{Habicht}, \citenamefont {Smeibidl},\ and\ \citenamefont
		{Kiefer}}]{coldea2010quantumcriticalityinanisingchain}%
	\BibitemOpen
	\bibfield  {author} {\bibinfo {author} {\bibfnamefont {R.}~\bibnamefont
			{Coldea}}, \bibinfo {author} {\bibfnamefont {D.~A.}\ \bibnamefont {Tennant}},
		\bibinfo {author} {\bibfnamefont {E.~M.}\ \bibnamefont {Wheeler}}, \bibinfo
		{author} {\bibfnamefont {E.}~\bibnamefont {Wawrzynska}}, \bibinfo {author}
		{\bibfnamefont {D.}~\bibnamefont {Prabhakaran}}, \bibinfo {author}
		{\bibfnamefont {M.}~\bibnamefont {Telling}}, \bibinfo {author} {\bibfnamefont
			{K.}~\bibnamefont {Habicht}}, \bibinfo {author} {\bibfnamefont
			{P.}~\bibnamefont {Smeibidl}}, \ and\ \bibinfo {author} {\bibfnamefont
			{K.}~\bibnamefont {Kiefer}},\ }\href {\doibase 10.1126/science.1180085}
	{\bibfield  {journal} {\bibinfo  {journal} {Science}\ }\textbf {\bibinfo
			{volume} {327}},\ \bibinfo {pages} {177} (\bibinfo {year}
		{2010})}\BibitemShut {NoStop}%
	\bibitem [{\citenamefont
		{Rutkevich}(2010)}]{rutkevich2010ontheweakconfinement}%
	\BibitemOpen
	\bibfield  {author} {\bibinfo {author} {\bibfnamefont {S.~B.}\ \bibnamefont
			{Rutkevich}},\ }\href {\doibase 10.1088/1742-5468/2010/07/P07015} {\bibfield
		{journal} {\bibinfo  {journal} {Journal of Statistical Mechanics: Theory and
				Experiment}\ }\textbf {\bibinfo {volume} {2010}},\ \bibinfo {pages} {P07015}
		(\bibinfo {year} {2010})}\BibitemShut {NoStop}%
	\bibitem [{\citenamefont {Kj\"all}\ \emph {et~al.}(2011)\citenamefont
		{Kj\"all}, \citenamefont {Pollmann},\ and\ \citenamefont
		{Moore}}]{kjaell2011boundstatesande8symmetry}%
	\BibitemOpen
	\bibfield  {author} {\bibinfo {author} {\bibfnamefont {J.~A.}\ \bibnamefont
			{Kj\"all}}, \bibinfo {author} {\bibfnamefont {F.}~\bibnamefont {Pollmann}}, \
		and\ \bibinfo {author} {\bibfnamefont {J.~E.}\ \bibnamefont {Moore}},\ }\href
	{\doibase 10.1103/PhysRevB.83.020407} {\bibfield  {journal} {\bibinfo
			{journal} {Phys. Rev. B}\ }\textbf {\bibinfo {volume} {83}},\ \bibinfo
		{pages} {020407} (\bibinfo {year} {2011})}\BibitemShut {NoStop}%
	\bibitem [{\citenamefont {Nandi}\ \emph {et~al.}(2019)\citenamefont {Nandi},
		\citenamefont {Prabhakaran},\ and\ \citenamefont
		{Mandal}}]{nandi2019spinchargelattice}%
	\BibitemOpen
	\bibfield  {author} {\bibinfo {author} {\bibfnamefont {M.}~\bibnamefont
			{Nandi}}, \bibinfo {author} {\bibfnamefont {D.}~\bibnamefont {Prabhakaran}},
		\ and\ \bibinfo {author} {\bibfnamefont {P.}~\bibnamefont {Mandal}},\ }\href
	{\doibase 10.1088/1361-648X/ab0539} {\bibfield  {journal} {\bibinfo
			{journal} {Journal of Physics: Condensed Matter}\ }\textbf {\bibinfo {volume}
			{31}},\ \bibinfo {pages} {195802} (\bibinfo {year} {2019})}\BibitemShut
	{NoStop}%
	\bibitem [{\citenamefont {Fava}\ \emph {et~al.}(2020)\citenamefont {Fava},
		\citenamefont {Coldea},\ and\ \citenamefont
		{Parameswaran}}]{fava2020glidesymmetrybreaking}%
	\BibitemOpen
	\bibfield  {author} {\bibinfo {author} {\bibfnamefont {M.}~\bibnamefont
			{Fava}}, \bibinfo {author} {\bibfnamefont {R.}~\bibnamefont {Coldea}}, \ and\
		\bibinfo {author} {\bibfnamefont {S.~A.}\ \bibnamefont {Parameswaran}},\
	}\href {\doibase 10.1073/pnas.2007986117} {\bibfield  {journal} {\bibinfo
			{journal} {Proceedings of the National Academy of Sciences}\ }\textbf
		{\bibinfo {volume} {117}},\ \bibinfo {pages} {25219} (\bibinfo {year}
		{2020})}\BibitemShut {NoStop}%
	\bibitem [{\citenamefont {Xu}\ \emph {et~al.}(2022)\citenamefont {Xu},
		\citenamefont {Wang}, \citenamefont {Huang}, \citenamefont {Ni},
		\citenamefont {Zhao}, \citenamefont {Dai}, \citenamefont {Pan}, \citenamefont
		{Hong}, \citenamefont {Chauhan}, \citenamefont {Koohpayeh}, \citenamefont
		{Armitage},\ and\ \citenamefont
		{Li}}]{xu2022quantumcriticalmagneticexcitations}%
	\BibitemOpen
	\bibfield  {author} {\bibinfo {author} {\bibfnamefont {Y.}~\bibnamefont
			{Xu}}, \bibinfo {author} {\bibfnamefont {L.~S.}\ \bibnamefont {Wang}},
		\bibinfo {author} {\bibfnamefont {Y.~Y.}\ \bibnamefont {Huang}}, \bibinfo
		{author} {\bibfnamefont {J.~M.}\ \bibnamefont {Ni}}, \bibinfo {author}
		{\bibfnamefont {C.~C.}\ \bibnamefont {Zhao}}, \bibinfo {author}
		{\bibfnamefont {Y.~F.}\ \bibnamefont {Dai}}, \bibinfo {author} {\bibfnamefont
			{B.~Y.}\ \bibnamefont {Pan}}, \bibinfo {author} {\bibfnamefont {X.~C.}\
			\bibnamefont {Hong}}, \bibinfo {author} {\bibfnamefont {P.}~\bibnamefont
			{Chauhan}}, \bibinfo {author} {\bibfnamefont {S.~M.}\ \bibnamefont
			{Koohpayeh}}, \bibinfo {author} {\bibfnamefont {N.~P.}\ \bibnamefont
			{Armitage}}, \ and\ \bibinfo {author} {\bibfnamefont {S.~Y.}\ \bibnamefont
			{Li}},\ }\href {\doibase 10.1103/PhysRevX.12.021020} {\bibfield  {journal}
		{\bibinfo  {journal} {Phys. Rev. X}\ }\textbf {\bibinfo {volume} {12}},\
		\bibinfo {pages} {021020} (\bibinfo {year} {2022})}\BibitemShut {NoStop}%
	\bibitem [{\citenamefont {Ringler}\ \emph {et~al.}(2022)\citenamefont
		{Ringler}, \citenamefont {Kolesnikov},\ and\ \citenamefont
		{Ross}}]{ringler2022singleionproperties}%
	\BibitemOpen
	\bibfield  {author} {\bibinfo {author} {\bibfnamefont {J.~A.}\ \bibnamefont
			{Ringler}}, \bibinfo {author} {\bibfnamefont {A.~I.}\ \bibnamefont
			{Kolesnikov}}, \ and\ \bibinfo {author} {\bibfnamefont {K.~A.}\ \bibnamefont
			{Ross}},\ }\href {\doibase 10.1103/PhysRevB.105.224421} {\bibfield  {journal}
		{\bibinfo  {journal} {Phys. Rev. B}\ }\textbf {\bibinfo {volume} {105}},\
		\bibinfo {pages} {224421} (\bibinfo {year} {2022})}\BibitemShut {NoStop}%
	\bibitem [{\citenamefont {Woodland}\ \emph
		{et~al.}(2023{\natexlab{a}})\citenamefont {Woodland}, \citenamefont {Lovas},
		\citenamefont {Telling}, \citenamefont {Prabhakaran}, \citenamefont
		{Balents},\ and\ \citenamefont
		{Coldea}}]{coldea2023excitationsofquantumisingchain}%
	\BibitemOpen
	\bibfield  {author} {\bibinfo {author} {\bibfnamefont {L.}~\bibnamefont
			{Woodland}}, \bibinfo {author} {\bibfnamefont {I.}~\bibnamefont {Lovas}},
		\bibinfo {author} {\bibfnamefont {M.}~\bibnamefont {Telling}}, \bibinfo
		{author} {\bibfnamefont {D.}~\bibnamefont {Prabhakaran}}, \bibinfo {author}
		{\bibfnamefont {L.}~\bibnamefont {Balents}}, \ and\ \bibinfo {author}
		{\bibfnamefont {R.}~\bibnamefont {Coldea}},\ }\href {\doibase
		10.1103/PhysRevB.108.184417} {\bibfield  {journal} {\bibinfo  {journal}
			{Phys. Rev. B}\ }\textbf {\bibinfo {volume} {108}},\ \bibinfo {pages}
		{184417} (\bibinfo {year} {2023}{\natexlab{a}})}\BibitemShut {NoStop}%
	\bibitem [{\citenamefont {Woodland}\ \emph
		{et~al.}(2023{\natexlab{b}})\citenamefont {Woodland}, \citenamefont
		{Macdougal}, \citenamefont {Cabrera}, \citenamefont {Thompson}, \citenamefont
		{Prabhakaran}, \citenamefont {Bewley},\ and\ \citenamefont
		{Coldea}}]{coldea2023tuning}%
	\BibitemOpen
	\bibfield  {author} {\bibinfo {author} {\bibfnamefont {L.}~\bibnamefont
			{Woodland}}, \bibinfo {author} {\bibfnamefont {D.}~\bibnamefont {Macdougal}},
		\bibinfo {author} {\bibfnamefont {I.~M.}\ \bibnamefont {Cabrera}}, \bibinfo
		{author} {\bibfnamefont {J.~D.}\ \bibnamefont {Thompson}}, \bibinfo {author}
		{\bibfnamefont {D.}~\bibnamefont {Prabhakaran}}, \bibinfo {author}
		{\bibfnamefont {R.~I.}\ \bibnamefont {Bewley}}, \ and\ \bibinfo {author}
		{\bibfnamefont {R.}~\bibnamefont {Coldea}},\ }\href {\doibase
		10.1103/PhysRevB.108.184416} {\bibfield  {journal} {\bibinfo  {journal}
			{Phys. Rev. B}\ }\textbf {\bibinfo {volume} {108}},\ \bibinfo {pages}
		{166002} (\bibinfo {year} {2023}{\natexlab{b}})}\BibitemShut {NoStop}%
	\bibitem [{\citenamefont {Liang}\ \emph {et~al.}(2015)\citenamefont {Liang},
		\citenamefont {Koohpayeh}, \citenamefont {Krizan}, \citenamefont {McQueen},
		\citenamefont {Cava},\ and\ \citenamefont {Ong}}]{liang2015heatcapacitypeak}%
	\BibitemOpen
	\bibfield  {author} {\bibinfo {author} {\bibfnamefont {T.}~\bibnamefont
			{Liang}}, \bibinfo {author} {\bibfnamefont {S.~M.}\ \bibnamefont
			{Koohpayeh}}, \bibinfo {author} {\bibfnamefont {J.~W.}\ \bibnamefont
			{Krizan}}, \bibinfo {author} {\bibfnamefont {T.~M.}\ \bibnamefont {McQueen}},
		\bibinfo {author} {\bibfnamefont {R.~J.}\ \bibnamefont {Cava}}, \ and\
		\bibinfo {author} {\bibfnamefont {N.~P.}\ \bibnamefont {Ong}},\ }\href@noop
	{} {\bibfield  {journal} {\bibinfo  {journal} {Nature Communications}\
		}\textbf {\bibinfo {volume} {6}} (\bibinfo {year} {2015})}\BibitemShut
	{NoStop}%
	\bibitem [{\citenamefont {Kinross}\ \emph {et~al.}(2014)\citenamefont
		{Kinross}, \citenamefont {Fu}, \citenamefont {Munsie}, \citenamefont
		{Dabkowska}, \citenamefont {Luke}, \citenamefont {Sachdev},\ and\
		\citenamefont {Imai}}]{kinross2014evolutionofquantumfluctuations}%
	\BibitemOpen
	\bibfield  {author} {\bibinfo {author} {\bibfnamefont {A.~W.}\ \bibnamefont
			{Kinross}}, \bibinfo {author} {\bibfnamefont {M.}~\bibnamefont {Fu}},
		\bibinfo {author} {\bibfnamefont {T.~J.}\ \bibnamefont {Munsie}}, \bibinfo
		{author} {\bibfnamefont {H.~A.}\ \bibnamefont {Dabkowska}}, \bibinfo {author}
		{\bibfnamefont {G.~M.}\ \bibnamefont {Luke}}, \bibinfo {author}
		{\bibfnamefont {S.}~\bibnamefont {Sachdev}}, \ and\ \bibinfo {author}
		{\bibfnamefont {T.}~\bibnamefont {Imai}},\ }\href {\doibase
		10.1103/PhysRevX.4.031008} {\bibfield  {journal} {\bibinfo  {journal} {Phys.
				Rev. X}\ }\textbf {\bibinfo {volume} {4}},\ \bibinfo {pages} {031008}
		(\bibinfo {year} {2014})}\BibitemShut {NoStop}%
	\bibitem [{\citenamefont {Morris}\ \emph {et~al.}(2014)\citenamefont {Morris},
		\citenamefont {Vald\'es~Aguilar}, \citenamefont {Ghosh}, \citenamefont
		{Koohpayeh}, \citenamefont {Krizan}, \citenamefont {Cava}, \citenamefont
		{Tchernyshyov}, \citenamefont {McQueen},\ and\ \citenamefont
		{Armitage}}]{morris2014hierarchyofboundstates}%
	\BibitemOpen
	\bibfield  {author} {\bibinfo {author} {\bibfnamefont {C.~M.}\ \bibnamefont
			{Morris}}, \bibinfo {author} {\bibfnamefont {R.}~\bibnamefont
			{Vald\'es~Aguilar}}, \bibinfo {author} {\bibfnamefont {A.}~\bibnamefont
			{Ghosh}}, \bibinfo {author} {\bibfnamefont {S.~M.}\ \bibnamefont
			{Koohpayeh}}, \bibinfo {author} {\bibfnamefont {J.}~\bibnamefont {Krizan}},
		\bibinfo {author} {\bibfnamefont {R.~J.}\ \bibnamefont {Cava}}, \bibinfo
		{author} {\bibfnamefont {O.}~\bibnamefont {Tchernyshyov}}, \bibinfo {author}
		{\bibfnamefont {T.~M.}\ \bibnamefont {McQueen}}, \ and\ \bibinfo {author}
		{\bibfnamefont {N.~P.}\ \bibnamefont {Armitage}},\ }\href {\doibase
		10.1103/PhysRevLett.112.137403} {\bibfield  {journal} {\bibinfo  {journal}
			{Phys. Rev. Lett.}\ }\textbf {\bibinfo {volume} {112}},\ \bibinfo {pages}
		{137403} (\bibinfo {year} {2014})}\BibitemShut {NoStop}%
	\bibitem [{\citenamefont {Amelin}\ \emph {et~al.}(2020)\citenamefont {Amelin},
		\citenamefont {Engelmayer}, \citenamefont {Viirok}, \citenamefont {Nagel},
		\citenamefont {R\~o\ om}, \citenamefont {Lorenz},\ and\ \citenamefont
		{Wang}}]{amelin2020experimentalobservation}%
	\BibitemOpen
	\bibfield  {author} {\bibinfo {author} {\bibfnamefont {K.}~\bibnamefont
			{Amelin}}, \bibinfo {author} {\bibfnamefont {J.}~\bibnamefont {Engelmayer}},
		\bibinfo {author} {\bibfnamefont {J.}~\bibnamefont {Viirok}}, \bibinfo
		{author} {\bibfnamefont {U.}~\bibnamefont {Nagel}}, \bibinfo {author}
		{\bibfnamefont {T.}~\bibnamefont {R\~o\ om}}, \bibinfo {author}
		{\bibfnamefont {T.}~\bibnamefont {Lorenz}}, \ and\ \bibinfo {author}
		{\bibfnamefont {Z.}~\bibnamefont {Wang}},\ }\href {\doibase
		10.1103/PhysRevB.102.104431} {\bibfield  {journal} {\bibinfo  {journal}
			{Phys. Rev. B}\ }\textbf {\bibinfo {volume} {102}},\ \bibinfo {pages}
		{104431} (\bibinfo {year} {2020})}\BibitemShut {NoStop}%
	\bibitem [{\citenamefont {Morris}\ \emph {et~al.}(2021)\citenamefont {Morris},
		\citenamefont {Desai}, \citenamefont {Viirok}, \citenamefont {Hüvonen},
		\citenamefont {Nagel}, \citenamefont {Rõõm}, \citenamefont {Krizan},
		\citenamefont {Cava}, \citenamefont {McQueen}, \citenamefont {Koohpayeh},
		\citenamefont {Kaul},\ and\ \citenamefont {Armitage}}]{armitage2021duality}%
	\BibitemOpen
	\bibfield  {author} {\bibinfo {author} {\bibfnamefont {C.~M.}\ \bibnamefont
			{Morris}}, \bibinfo {author} {\bibfnamefont {N.}~\bibnamefont {Desai}},
		\bibinfo {author} {\bibfnamefont {J.}~\bibnamefont {Viirok}}, \bibinfo
		{author} {\bibfnamefont {D.}~\bibnamefont {Hüvonen}}, \bibinfo {author}
		{\bibfnamefont {U.}~\bibnamefont {Nagel}}, \bibinfo {author} {\bibfnamefont
			{T.}~\bibnamefont {Rõõm}}, \bibinfo {author} {\bibfnamefont {J.~W.}\
			\bibnamefont {Krizan}}, \bibinfo {author} {\bibfnamefont {R.~J.}\
			\bibnamefont {Cava}}, \bibinfo {author} {\bibfnamefont {T.~M.}\ \bibnamefont
			{McQueen}}, \bibinfo {author} {\bibfnamefont {S.~M.}\ \bibnamefont
			{Koohpayeh}}, \bibinfo {author} {\bibfnamefont {R.~K.}\ \bibnamefont {Kaul}},
		\ and\ \bibinfo {author} {\bibfnamefont {N.~P.}\ \bibnamefont {Armitage}},\
	}\href {\doibase 10.1038/s41567-021-01208-0} {\bibfield  {journal} {\bibinfo
			{journal} {Nature Physics}\ }\textbf {\bibinfo {volume} {17}},\ \bibinfo
		{pages} {832} (\bibinfo {year} {2021})}\BibitemShut {NoStop}%
	\bibitem [{\citenamefont {Amelin}\ \emph {et~al.}(2022)\citenamefont {Amelin},
		\citenamefont {Viirok}, \citenamefont {Nagel}, \citenamefont {Rõõm},
		\citenamefont {Engelmayer}, \citenamefont {Dey}, \citenamefont {Nugroho},
		\citenamefont {Lorenz},\ and\ \citenamefont
		{Wang}}]{amelin2022quantumspindynamics}%
	\BibitemOpen
	\bibfield  {author} {\bibinfo {author} {\bibfnamefont {K.}~\bibnamefont
			{Amelin}}, \bibinfo {author} {\bibfnamefont {J.}~\bibnamefont {Viirok}},
		\bibinfo {author} {\bibfnamefont {U.}~\bibnamefont {Nagel}}, \bibinfo
		{author} {\bibfnamefont {T.}~\bibnamefont {Rõõm}}, \bibinfo {author}
		{\bibfnamefont {J.}~\bibnamefont {Engelmayer}}, \bibinfo {author}
		{\bibfnamefont {T.}~\bibnamefont {Dey}}, \bibinfo {author} {\bibfnamefont
			{A.~A.}\ \bibnamefont {Nugroho}}, \bibinfo {author} {\bibfnamefont
			{T.}~\bibnamefont {Lorenz}}, \ and\ \bibinfo {author} {\bibfnamefont
			{Z.}~\bibnamefont {Wang}},\ }\href {\doibase 10.1088/1751-8121/aca6b8}
	{\bibfield  {journal} {\bibinfo  {journal} {Journal of Physics A:
				Mathematical and Theoretical}\ }\textbf {\bibinfo {volume} {55}},\ \bibinfo
		{pages} {484005} (\bibinfo {year} {2022})}\BibitemShut {NoStop}%
	\bibitem [{\citenamefont {Lines}(1963)}]{lines1963magneticproperties}%
	\BibitemOpen
	\bibfield  {author} {\bibinfo {author} {\bibfnamefont {M.~E.}\ \bibnamefont
			{Lines}},\ }\href {\doibase 10.1103/PhysRev.131.546} {\bibfield  {journal}
		{\bibinfo  {journal} {Phys. Rev.}\ }\textbf {\bibinfo {volume} {131}},\
		\bibinfo {pages} {546} (\bibinfo {year} {1963})}\BibitemShut {NoStop}%
	\bibitem [{\citenamefont {Oguchi}(1965)}]{oguchi1965theoryofmagnetism}%
	\BibitemOpen
	\bibfield  {author} {\bibinfo {author} {\bibfnamefont {T.}~\bibnamefont
			{Oguchi}},\ }\href {\doibase 10.1143/JPSJ.20.2236} {\bibfield  {journal}
		{\bibinfo  {journal} {Journal of the Physical Society of Japan}\ }\textbf
		{\bibinfo {volume} {20}},\ \bibinfo {pages} {2236} (\bibinfo {year}
		{1965})}\BibitemShut {NoStop}%
	\bibitem [{\citenamefont {You}\ \emph {et~al.}(2014)\citenamefont {You},
		\citenamefont {Liu}, \citenamefont {Horsch},\ and\ \citenamefont
		{Ole{\'s}}}]{you2014exact}%
	\BibitemOpen
	\bibfield  {author} {\bibinfo {author} {\bibfnamefont {W.-L.}\ \bibnamefont
			{You}}, \bibinfo {author} {\bibfnamefont {G.-H.}\ \bibnamefont {Liu}},
		\bibinfo {author} {\bibfnamefont {P.}~\bibnamefont {Horsch}}, \ and\ \bibinfo
		{author} {\bibfnamefont {A.~M.}\ \bibnamefont {Ole{\'s}}},\ }\href@noop {}
	{\bibfield  {journal} {\bibinfo  {journal} {Physical Review B}\ }\textbf
		{\bibinfo {volume} {90}},\ \bibinfo {pages} {094413} (\bibinfo {year}
		{2014})}\BibitemShut {NoStop}%
	\bibitem [{\citenamefont {Liu}\ and\ \citenamefont
		{Khaliullin}(2018)}]{liu2018pseudospin}%
	\BibitemOpen
	\bibfield  {author} {\bibinfo {author} {\bibfnamefont {H.}~\bibnamefont
			{Liu}}\ and\ \bibinfo {author} {\bibfnamefont {G.}~\bibnamefont
			{Khaliullin}},\ }\href {\doibase 10.1103/PhysRevB.97.014407} {\bibfield
		{journal} {\bibinfo  {journal} {Phys. Rev. B}\ }\textbf {\bibinfo {volume}
			{97}},\ \bibinfo {pages} {014407} (\bibinfo {year} {2018})}\BibitemShut
	{NoStop}%
	\bibitem [{\citenamefont {Liu}\ \emph {et~al.}(2020{\natexlab{a}})\citenamefont
		{Liu}, \citenamefont {Chaloupka},\ and\ \citenamefont
		{Khaliullin}}]{liu2020kitaevspinliquid}%
	\BibitemOpen
	\bibfield  {author} {\bibinfo {author} {\bibfnamefont {H.}~\bibnamefont
			{Liu}}, \bibinfo {author} {\bibfnamefont {J.~c.~v.}\ \bibnamefont
			{Chaloupka}}, \ and\ \bibinfo {author} {\bibfnamefont {G.}~\bibnamefont
			{Khaliullin}},\ }\href {\doibase 10.1103/PhysRevLett.125.047201} {\bibfield
		{journal} {\bibinfo  {journal} {Phys. Rev. Lett.}\ }\textbf {\bibinfo
			{volume} {125}},\ \bibinfo {pages} {047201} (\bibinfo {year}
		{2020}{\natexlab{a}})}\BibitemShut {NoStop}%
	\bibitem [{\citenamefont {Liu}\ \emph {et~al.}(2020{\natexlab{b}})\citenamefont
		{Liu}, \citenamefont {Chaloupka},\ and\ \citenamefont
		{Khaliullin}}]{liu2021towardskitaevspinliquid}%
	\BibitemOpen
	\bibfield  {author} {\bibinfo {author} {\bibfnamefont {H.}~\bibnamefont
			{Liu}}, \bibinfo {author} {\bibfnamefont {J.~c.~v.}\ \bibnamefont
			{Chaloupka}}, \ and\ \bibinfo {author} {\bibfnamefont {G.}~\bibnamefont
			{Khaliullin}},\ }\href {\doibase 10.1103/PhysRevLett.125.047201} {\bibfield
		{journal} {\bibinfo  {journal} {Phys. Rev. Lett.}\ }\textbf {\bibinfo
			{volume} {125}},\ \bibinfo {pages} {047201} (\bibinfo {year}
		{2020}{\natexlab{b}})}\BibitemShut {NoStop}%
	\bibitem [{\citenamefont {Sano}\ \emph {et~al.}(2018)\citenamefont {Sano},
		\citenamefont {Kato},\ and\ \citenamefont
		{Motome}}]{sano2018kitaevheisenberghamiltonian}%
	\BibitemOpen
	\bibfield  {author} {\bibinfo {author} {\bibfnamefont {R.}~\bibnamefont
			{Sano}}, \bibinfo {author} {\bibfnamefont {Y.}~\bibnamefont {Kato}}, \ and\
		\bibinfo {author} {\bibfnamefont {Y.}~\bibnamefont {Motome}},\ }\href
	{\doibase 10.1103/PhysRevB.97.014408} {\bibfield  {journal} {\bibinfo
			{journal} {Phys. Rev. B}\ }\textbf {\bibinfo {volume} {97}},\ \bibinfo
		{pages} {014408} (\bibinfo {year} {2018})}\BibitemShut {NoStop}%
	\bibitem [{\citenamefont {Kitaev}(2006)}]{kitaev2006anyons}%
	\BibitemOpen
	\bibfield  {author} {\bibinfo {author} {\bibfnamefont {A.}~\bibnamefont
			{Kitaev}},\ }\href {\doibase https://doi.org/10.1016/j.aop.2005.10.005}
	{\bibfield  {journal} {\bibinfo  {journal} {Annals of Physics}\ }\textbf
		{\bibinfo {volume} {321}},\ \bibinfo {pages} {2} (\bibinfo {year} {2006})},\
	\bibinfo {note} {january Special Issue}\BibitemShut {NoStop}%
	\bibitem [{\citenamefont {Jackeli}\ and\ \citenamefont
		{Khaliullin}(2009)}]{jackeli2009}%
	\BibitemOpen
	\bibfield  {author} {\bibinfo {author} {\bibfnamefont {G.}~\bibnamefont
			{Jackeli}}\ and\ \bibinfo {author} {\bibfnamefont {G.}~\bibnamefont
			{Khaliullin}},\ }\href@noop {} {\bibfield  {journal} {\bibinfo  {journal}
			{Physical review letters}\ }\textbf {\bibinfo {volume} {102}},\ \bibinfo
		{pages} {017205} (\bibinfo {year} {2009})}\BibitemShut {NoStop}%
	\bibitem [{\citenamefont {Rau}\ \emph {et~al.}(2014)\citenamefont {Rau},
		\citenamefont {Lee},\ and\ \citenamefont
		{Kee}}]{haeyoung2014genericspinmodel}%
	\BibitemOpen
	\bibfield  {author} {\bibinfo {author} {\bibfnamefont {J.~G.}\ \bibnamefont
			{Rau}}, \bibinfo {author} {\bibfnamefont {E.~K.-H.}\ \bibnamefont {Lee}}, \
		and\ \bibinfo {author} {\bibfnamefont {H.-Y.}\ \bibnamefont {Kee}},\ }\href
	{\doibase 10.1103/PhysRevLett.112.077204} {\bibfield  {journal} {\bibinfo
			{journal} {Phys. Rev. Lett.}\ }\textbf {\bibinfo {volume} {112}},\ \bibinfo
		{pages} {077204} (\bibinfo {year} {2014})}\BibitemShut {NoStop}%
	\bibitem [{\citenamefont {Winter}\ \emph {et~al.}(2016)\citenamefont {Winter},
		\citenamefont {Li}, \citenamefont {Jeschke},\ and\ \citenamefont
		{Valent{\'\i}}}]{winter2016challenges}%
	\BibitemOpen
	\bibfield  {author} {\bibinfo {author} {\bibfnamefont {S.~M.}\ \bibnamefont
			{Winter}}, \bibinfo {author} {\bibfnamefont {Y.}~\bibnamefont {Li}}, \bibinfo
		{author} {\bibfnamefont {H.~O.}\ \bibnamefont {Jeschke}}, \ and\ \bibinfo
		{author} {\bibfnamefont {R.}~\bibnamefont {Valent{\'\i}}},\ }\href@noop {}
	{\bibfield  {journal} {\bibinfo  {journal} {Physical Review B}\ }\textbf
		{\bibinfo {volume} {93}},\ \bibinfo {pages} {214431} (\bibinfo {year}
		{2016})}\BibitemShut {NoStop}%
	\bibitem [{\citenamefont {Winter}\ \emph
		{et~al.}(2017{\natexlab{a}})\citenamefont {Winter}, \citenamefont {Tsirlin},
		\citenamefont {Daghofer}, \citenamefont {van~den Brink}, \citenamefont
		{Singh}, \citenamefont {Gegenwart},\ and\ \citenamefont
		{Valent{\'\i}}}]{winter2017models}%
	\BibitemOpen
	\bibfield  {author} {\bibinfo {author} {\bibfnamefont {S.~M.}\ \bibnamefont
			{Winter}}, \bibinfo {author} {\bibfnamefont {A.~A.}\ \bibnamefont {Tsirlin}},
		\bibinfo {author} {\bibfnamefont {M.}~\bibnamefont {Daghofer}}, \bibinfo
		{author} {\bibfnamefont {J.}~\bibnamefont {van~den Brink}}, \bibinfo {author}
		{\bibfnamefont {Y.}~\bibnamefont {Singh}}, \bibinfo {author} {\bibfnamefont
			{P.}~\bibnamefont {Gegenwart}}, \ and\ \bibinfo {author} {\bibfnamefont
			{R.}~\bibnamefont {Valent{\'\i}}},\ }\href@noop {} {\bibfield  {journal}
		{\bibinfo  {journal} {Journal of Physics: Condensed Matter}\ }\textbf
		{\bibinfo {volume} {29}},\ \bibinfo {pages} {493002} (\bibinfo {year}
		{2017}{\natexlab{a}})}\BibitemShut {NoStop}%
	\bibitem [{\citenamefont {Maksimov}\ \emph {et~al.}(2019)\citenamefont
		{Maksimov}, \citenamefont {Zhu}, \citenamefont {White},\ and\ \citenamefont
		{Chernyshev}}]{maksimov2019anisotropicexchange}%
	\BibitemOpen
	\bibfield  {author} {\bibinfo {author} {\bibfnamefont {P.~A.}\ \bibnamefont
			{Maksimov}}, \bibinfo {author} {\bibfnamefont {Z.}~\bibnamefont {Zhu}},
		\bibinfo {author} {\bibfnamefont {S.~R.}\ \bibnamefont {White}}, \ and\
		\bibinfo {author} {\bibfnamefont {A.~L.}\ \bibnamefont {Chernyshev}},\ }\href
	{\doibase 10.1103/PhysRevX.9.021017} {\bibfield  {journal} {\bibinfo
			{journal} {Phys. Rev. X}\ }\textbf {\bibinfo {volume} {9}},\ \bibinfo {pages}
		{021017} (\bibinfo {year} {2019})}\BibitemShut {NoStop}%
	\bibitem [{\citenamefont {Takagi}\ \emph {et~al.}(2019)\citenamefont {Takagi},
		\citenamefont {Takayama}, \citenamefont {Jackeli}, \citenamefont
		{Khaliullin},\ and\ \citenamefont {Nagler}}]{takagi2019}%
	\BibitemOpen
	\bibfield  {author} {\bibinfo {author} {\bibfnamefont {H.}~\bibnamefont
			{Takagi}}, \bibinfo {author} {\bibfnamefont {T.}~\bibnamefont {Takayama}},
		\bibinfo {author} {\bibfnamefont {G.}~\bibnamefont {Jackeli}}, \bibinfo
		{author} {\bibfnamefont {G.}~\bibnamefont {Khaliullin}}, \ and\ \bibinfo
		{author} {\bibfnamefont {S.~E.}\ \bibnamefont {Nagler}},\ }\href@noop {}
	{\bibfield  {journal} {\bibinfo  {journal} {Nature Reviews Physics}\ }\textbf
		{\bibinfo {volume} {1}},\ \bibinfo {pages} {264} (\bibinfo {year}
		{2019})}\BibitemShut {NoStop}%
	\bibitem [{\citenamefont {Trebst}\ and\ \citenamefont
		{Hickey}(2022)}]{trebst2022kitaev}%
	\BibitemOpen
	\bibfield  {author} {\bibinfo {author} {\bibfnamefont {S.}~\bibnamefont
			{Trebst}}\ and\ \bibinfo {author} {\bibfnamefont {C.}~\bibnamefont
			{Hickey}},\ }\href@noop {} {\bibfield  {journal} {\bibinfo  {journal}
			{Physics Reports}\ }\textbf {\bibinfo {volume} {950}},\ \bibinfo {pages} {1}
		(\bibinfo {year} {2022})}\BibitemShut {NoStop}%
	\bibitem [{\citenamefont {Das}\ \emph {et~al.}(2021)\citenamefont {Das},
		\citenamefont {Voleti}, \citenamefont {Saha-Dasgupta},\ and\ \citenamefont
		{Paramekanti}}]{das2021xy}%
	\BibitemOpen
	\bibfield  {author} {\bibinfo {author} {\bibfnamefont {S.}~\bibnamefont
			{Das}}, \bibinfo {author} {\bibfnamefont {S.}~\bibnamefont {Voleti}},
		\bibinfo {author} {\bibfnamefont {T.}~\bibnamefont {Saha-Dasgupta}}, \ and\
		\bibinfo {author} {\bibfnamefont {A.}~\bibnamefont {Paramekanti}},\
	}\href@noop {} {\bibfield  {journal} {\bibinfo  {journal} {Physical Review
				B}\ }\textbf {\bibinfo {volume} {104}},\ \bibinfo {pages} {134425} (\bibinfo
		{year} {2021})}\BibitemShut {NoStop}%
	\bibitem [{\citenamefont {Chen}\ \emph {et~al.}(2021)\citenamefont {Chen},
		\citenamefont {Li}, \citenamefont {Hu}, \citenamefont {Hu}, \citenamefont
		{Yue}, \citenamefont {Sutarto}, \citenamefont {He}, \citenamefont {Iida},
		\citenamefont {Kamazawa}, \citenamefont {Yu} \emph {et~al.}}]{chen2021spin}%
	\BibitemOpen
	\bibfield  {author} {\bibinfo {author} {\bibfnamefont {W.}~\bibnamefont
			{Chen}}, \bibinfo {author} {\bibfnamefont {X.}~\bibnamefont {Li}}, \bibinfo
		{author} {\bibfnamefont {Z.}~\bibnamefont {Hu}}, \bibinfo {author}
		{\bibfnamefont {Z.}~\bibnamefont {Hu}}, \bibinfo {author} {\bibfnamefont
			{L.}~\bibnamefont {Yue}}, \bibinfo {author} {\bibfnamefont {R.}~\bibnamefont
			{Sutarto}}, \bibinfo {author} {\bibfnamefont {F.}~\bibnamefont {He}},
		\bibinfo {author} {\bibfnamefont {K.}~\bibnamefont {Iida}}, \bibinfo {author}
		{\bibfnamefont {K.}~\bibnamefont {Kamazawa}}, \bibinfo {author}
		{\bibfnamefont {W.}~\bibnamefont {Yu}},  \emph {et~al.},\ }\href@noop {}
	{\bibfield  {journal} {\bibinfo  {journal} {Physical Review B}\ }\textbf
		{\bibinfo {volume} {103}},\ \bibinfo {pages} {L180404} (\bibinfo {year}
		{2021})}\BibitemShut {NoStop}%
	\bibitem [{\citenamefont {Sanders}\ \emph {et~al.}(2022)\citenamefont
		{Sanders}, \citenamefont {Mole}, \citenamefont {Liu}, \citenamefont {Brown},
		\citenamefont {Yu}, \citenamefont {Ling},\ and\ \citenamefont
		{Rachel}}]{sanders2022dominant}%
	\BibitemOpen
	\bibfield  {author} {\bibinfo {author} {\bibfnamefont {A.~L.}\ \bibnamefont
			{Sanders}}, \bibinfo {author} {\bibfnamefont {R.~A.}\ \bibnamefont {Mole}},
		\bibinfo {author} {\bibfnamefont {J.}~\bibnamefont {Liu}}, \bibinfo {author}
		{\bibfnamefont {A.~J.}\ \bibnamefont {Brown}}, \bibinfo {author}
		{\bibfnamefont {D.}~\bibnamefont {Yu}}, \bibinfo {author} {\bibfnamefont
			{C.~D.}\ \bibnamefont {Ling}}, \ and\ \bibinfo {author} {\bibfnamefont
			{S.}~\bibnamefont {Rachel}},\ }\href@noop {} {\bibfield  {journal} {\bibinfo
			{journal} {Physical Review B}\ }\textbf {\bibinfo {volume} {106}},\ \bibinfo
		{pages} {014413} (\bibinfo {year} {2022})}\BibitemShut {NoStop}%
	\bibitem [{\citenamefont {Winter}(2022)}]{winter2022magneticcouplings}%
	\BibitemOpen
	\bibfield  {author} {\bibinfo {author} {\bibfnamefont {S.~M.}\ \bibnamefont
			{Winter}},\ }\href {\doibase 10.1088/2515-7639/ac94f8} {\bibfield  {journal}
		{\bibinfo  {journal} {J. Phys. Mater.}\ }\textbf {\bibinfo {volume} {5}},\
		\bibinfo {pages} {045003} (\bibinfo {year} {2022})}\BibitemShut {NoStop}%
	\bibitem [{\citenamefont {Zhang}\ \emph {et~al.}(2023)\citenamefont {Zhang},
		\citenamefont {Xu}, \citenamefont {Halloran}, \citenamefont {Zhong},
		\citenamefont {Broholm}, \citenamefont {Cava}, \citenamefont {Drichko},\ and\
		\citenamefont {Armitage}}]{zhang2023magnetic}%
	\BibitemOpen
	\bibfield  {author} {\bibinfo {author} {\bibfnamefont {X.}~\bibnamefont
			{Zhang}}, \bibinfo {author} {\bibfnamefont {Y.}~\bibnamefont {Xu}}, \bibinfo
		{author} {\bibfnamefont {T.}~\bibnamefont {Halloran}}, \bibinfo {author}
		{\bibfnamefont {R.}~\bibnamefont {Zhong}}, \bibinfo {author} {\bibfnamefont
			{C.}~\bibnamefont {Broholm}}, \bibinfo {author} {\bibfnamefont
			{R.}~\bibnamefont {Cava}}, \bibinfo {author} {\bibfnamefont {N.}~\bibnamefont
			{Drichko}}, \ and\ \bibinfo {author} {\bibfnamefont {N.}~\bibnamefont
			{Armitage}},\ }\href@noop {} {\bibfield  {journal} {\bibinfo  {journal}
			{Nature Materials}\ }\textbf {\bibinfo {volume} {22}},\ \bibinfo {pages} {58}
		(\bibinfo {year} {2023})}\BibitemShut {NoStop}%
	\bibitem [{\citenamefont {Liu}\ and\ \citenamefont {Kee}(2023)}]{liu2023non}%
	\BibitemOpen
	\bibfield  {author} {\bibinfo {author} {\bibfnamefont {X.}~\bibnamefont
			{Liu}}\ and\ \bibinfo {author} {\bibfnamefont {H.-Y.}\ \bibnamefont {Kee}},\
	}\href@noop {} {\bibfield  {journal} {\bibinfo  {journal} {Physical Review
				B}\ }\textbf {\bibinfo {volume} {107}},\ \bibinfo {pages} {054420} (\bibinfo
		{year} {2023})}\BibitemShut {NoStop}%
	\bibitem [{\citenamefont {Halloran}\ \emph {et~al.}(2023)\citenamefont
		{Halloran}, \citenamefont {Desrochers}, \citenamefont {Zhang}, \citenamefont
		{Chen}, \citenamefont {Chern}, \citenamefont {Xu}, \citenamefont {Winn},
		\citenamefont {Graves-Brook}, \citenamefont {Stone}, \citenamefont
		{Kolesnikov} \emph {et~al.}}]{halloran2023geometrical}%
	\BibitemOpen
	\bibfield  {author} {\bibinfo {author} {\bibfnamefont {T.}~\bibnamefont
			{Halloran}}, \bibinfo {author} {\bibfnamefont {F.}~\bibnamefont
			{Desrochers}}, \bibinfo {author} {\bibfnamefont {E.~Z.}\ \bibnamefont
			{Zhang}}, \bibinfo {author} {\bibfnamefont {T.}~\bibnamefont {Chen}},
		\bibinfo {author} {\bibfnamefont {L.~E.}\ \bibnamefont {Chern}}, \bibinfo
		{author} {\bibfnamefont {Z.}~\bibnamefont {Xu}}, \bibinfo {author}
		{\bibfnamefont {B.}~\bibnamefont {Winn}}, \bibinfo {author} {\bibfnamefont
			{M.}~\bibnamefont {Graves-Brook}}, \bibinfo {author} {\bibfnamefont
			{M.}~\bibnamefont {Stone}}, \bibinfo {author} {\bibfnamefont {A.~I.}\
			\bibnamefont {Kolesnikov}},  \emph {et~al.},\ }\href@noop {} {\bibfield
		{journal} {\bibinfo  {journal} {Proceedings of the National Academy of
				Sciences}\ }\textbf {\bibinfo {volume} {120}},\ \bibinfo {pages}
		{e2215509119} (\bibinfo {year} {2023})}\BibitemShut {NoStop}%
	\bibitem [{\citenamefont {Xiang}\ \emph {et~al.}(2023)\citenamefont {Xiang},
		\citenamefont {Dhakal}, \citenamefont {Ozerov}, \citenamefont {Jiang},
		\citenamefont {Mou}, \citenamefont {Ozarowski}, \citenamefont {Huang},
		\citenamefont {Zhou}, \citenamefont {Fang}, \citenamefont {Winter} \emph
		{et~al.}}]{xiang2023disorder}%
	\BibitemOpen
	\bibfield  {author} {\bibinfo {author} {\bibfnamefont {L.}~\bibnamefont
			{Xiang}}, \bibinfo {author} {\bibfnamefont {R.}~\bibnamefont {Dhakal}},
		\bibinfo {author} {\bibfnamefont {M.}~\bibnamefont {Ozerov}}, \bibinfo
		{author} {\bibfnamefont {Y.}~\bibnamefont {Jiang}}, \bibinfo {author}
		{\bibfnamefont {B.~S.}\ \bibnamefont {Mou}}, \bibinfo {author} {\bibfnamefont
			{A.}~\bibnamefont {Ozarowski}}, \bibinfo {author} {\bibfnamefont
			{Q.}~\bibnamefont {Huang}}, \bibinfo {author} {\bibfnamefont
			{H.}~\bibnamefont {Zhou}}, \bibinfo {author} {\bibfnamefont {J.}~\bibnamefont
			{Fang}}, \bibinfo {author} {\bibfnamefont {S.~M.}\ \bibnamefont {Winter}},
		\emph {et~al.},\ }\href@noop {} {\bibfield  {journal} {\bibinfo  {journal}
			{Physical Review Letters}\ }\textbf {\bibinfo {volume} {131}},\ \bibinfo
		{pages} {076701} (\bibinfo {year} {2023})}\BibitemShut {NoStop}%
	\bibitem [{\citenamefont {Bhattacharyya}\ \emph {et~al.}(2024)\citenamefont
		{Bhattacharyya}, \citenamefont {Petersen}, \citenamefont {Nishimoto},\ and\
		\citenamefont {Hozoi}}]{bhattacharyya2024kitaev}%
	\BibitemOpen
	\bibfield  {author} {\bibinfo {author} {\bibfnamefont {P.}~\bibnamefont
			{Bhattacharyya}}, \bibinfo {author} {\bibfnamefont {T.}~\bibnamefont
			{Petersen}}, \bibinfo {author} {\bibfnamefont {S.}~\bibnamefont {Nishimoto}},
		\ and\ \bibinfo {author} {\bibfnamefont {L.}~\bibnamefont {Hozoi}},\
	}\href@noop {} {\bibfield  {journal} {\bibinfo  {journal} {arXiv preprint
				arXiv:2404.12742}\ } (\bibinfo {year} {2024})}\BibitemShut {NoStop}%
	\bibitem [{\citenamefont {Churchill}\ and\ \citenamefont
		{Kee}(2024)}]{churchill2024transforming}%
	\BibitemOpen
	\bibfield  {author} {\bibinfo {author} {\bibfnamefont {D.}~\bibnamefont
			{Churchill}}\ and\ \bibinfo {author} {\bibfnamefont {H.-Y.}\ \bibnamefont
			{Kee}},\ }\href@noop {} {\  (\bibinfo {year} {2024})},\ \Eprint
	{http://arxiv.org/abs/2403.14754} {arXiv:2403.14754 [cond-mat.str-el]}
	\BibitemShut {NoStop}%
	\bibitem [{\citenamefont {Gallegos}\ and\ \citenamefont
		{Chernyshev}(2024)}]{gallegos2024MagnonInteractionsQuantum}%
	\BibitemOpen
	\bibfield  {author} {\bibinfo {author} {\bibfnamefont {C.~A.}\ \bibnamefont
			{Gallegos}}\ and\ \bibinfo {author} {\bibfnamefont {A.~L.}\ \bibnamefont
			{Chernyshev}},\ }\href {\doibase 10.1103/PhysRevB.109.014424} {\bibfield
		{journal} {\bibinfo  {journal} {Physical Review B}\ }\textbf {\bibinfo
			{volume} {109}},\ \bibinfo {pages} {014424} (\bibinfo {year}
		{2024})}\BibitemShut {NoStop}%
	\bibitem [{\citenamefont {Kobayashi}\ \emph {et~al.}(1999)\citenamefont
		{Kobayashi}, \citenamefont {Mitsuda}, \citenamefont {Ishikawa}, \citenamefont
		{Miyatani},\ and\ \citenamefont {Kohn}}]{kohn1999threedimensional}%
	\BibitemOpen
	\bibfield  {author} {\bibinfo {author} {\bibfnamefont {S.}~\bibnamefont
			{Kobayashi}}, \bibinfo {author} {\bibfnamefont {S.}~\bibnamefont {Mitsuda}},
		\bibinfo {author} {\bibfnamefont {M.}~\bibnamefont {Ishikawa}}, \bibinfo
		{author} {\bibfnamefont {K.}~\bibnamefont {Miyatani}}, \ and\ \bibinfo
		{author} {\bibfnamefont {K.}~\bibnamefont {Kohn}},\ }\href {\doibase
		10.1103/PhysRevB.60.3331} {\bibfield  {journal} {\bibinfo  {journal} {Phys.
				Rev. B}\ }\textbf {\bibinfo {volume} {60}},\ \bibinfo {pages} {3331}
		(\bibinfo {year} {1999})}\BibitemShut {NoStop}%
	\bibitem [{\citenamefont {Scharf}\ \emph
		{et~al.}(1979{\natexlab{b}})\citenamefont {Scharf}, \citenamefont {Weitzel},
		\citenamefont {Yaeger}, \citenamefont {Maartense},\ and\ \citenamefont
		{Wanklyn}}]{wanklyn1979magneticstructures}%
	\BibitemOpen
	\bibfield  {author} {\bibinfo {author} {\bibfnamefont {W.}~\bibnamefont
			{Scharf}}, \bibinfo {author} {\bibfnamefont {H.}~\bibnamefont {Weitzel}},
		\bibinfo {author} {\bibfnamefont {I.}~\bibnamefont {Yaeger}}, \bibinfo
		{author} {\bibfnamefont {I.}~\bibnamefont {Maartense}}, \ and\ \bibinfo
		{author} {\bibfnamefont {B.}~\bibnamefont {Wanklyn}},\ }\href {\doibase
		https://doi.org/10.1016/0304-8853(79)90044-1} {\bibfield  {journal} {\bibinfo
			{journal} {Journal of Magnetism and Magnetic Materials}\ }\textbf {\bibinfo
			{volume} {13}},\ \bibinfo {pages} {121} (\bibinfo {year}
		{1979}{\natexlab{b}})}\BibitemShut {NoStop}%
	\bibitem [{\citenamefont {Mitsuda}\ \emph {et~al.}(1994)\citenamefont
		{Mitsuda}, \citenamefont {Hosoya}, \citenamefont {Wada}, \citenamefont
		{Yoshizawa}, \citenamefont {Hanawa}, \citenamefont {Ishikawa}, \citenamefont
		{Miyatani}, \citenamefont {Saito},\ and\ \citenamefont
		{Kohn}}]{SetsuoMitsuda1994magordering}%
	\BibitemOpen
	\bibfield  {author} {\bibinfo {author} {\bibfnamefont {S.}~\bibnamefont
			{Mitsuda}}, \bibinfo {author} {\bibfnamefont {K.}~\bibnamefont {Hosoya}},
		\bibinfo {author} {\bibfnamefont {T.}~\bibnamefont {Wada}}, \bibinfo {author}
		{\bibfnamefont {H.}~\bibnamefont {Yoshizawa}}, \bibinfo {author}
		{\bibfnamefont {T.}~\bibnamefont {Hanawa}}, \bibinfo {author} {\bibfnamefont
			{M.}~\bibnamefont {Ishikawa}}, \bibinfo {author} {\bibfnamefont
			{K.}~\bibnamefont {Miyatani}}, \bibinfo {author} {\bibfnamefont
			{K.}~\bibnamefont {Saito}}, \ and\ \bibinfo {author} {\bibfnamefont
			{K.}~\bibnamefont {Kohn}},\ }\href {\doibase 10.1143/jpsj.63.3568} {\bibfield
		{journal} {\bibinfo  {journal} {Journal of the Physical Society of Japan}\
		}\textbf {\bibinfo {volume} {63}},\ \bibinfo {pages} {3568} (\bibinfo {year}
		{1994})}\BibitemShut {NoStop}%
	\bibitem [{\citenamefont {Riedl}\ \emph {et~al.}(2019)\citenamefont {Riedl},
		\citenamefont {Li}, \citenamefont {Valent{\'\i}},\ and\ \citenamefont
		{Winter}}]{riedl2019}%
	\BibitemOpen
	\bibfield  {author} {\bibinfo {author} {\bibfnamefont {K.}~\bibnamefont
			{Riedl}}, \bibinfo {author} {\bibfnamefont {Y.}~\bibnamefont {Li}}, \bibinfo
		{author} {\bibfnamefont {R.}~\bibnamefont {Valent{\'\i}}}, \ and\ \bibinfo
		{author} {\bibfnamefont {S.~M.}\ \bibnamefont {Winter}},\ }\href@noop {}
	{\bibfield  {journal} {\bibinfo  {journal} {physica status solidi (b)}\
		}\textbf {\bibinfo {volume} {256}},\ \bibinfo {pages} {1800684} (\bibinfo
		{year} {2019})}\BibitemShut {NoStop}%
	\bibitem [{\citenamefont {Winter}\ \emph
		{et~al.}(2017{\natexlab{b}})\citenamefont {Winter}, \citenamefont {Riedl},
		\citenamefont {Maksimov}, \citenamefont {Chernyshev}, \citenamefont
		{Honecker},\ and\ \citenamefont {Valent{\'\i}}}]{winter2017breakdown}%
	\BibitemOpen
	\bibfield  {author} {\bibinfo {author} {\bibfnamefont {S.~M.}\ \bibnamefont
			{Winter}}, \bibinfo {author} {\bibfnamefont {K.}~\bibnamefont {Riedl}},
		\bibinfo {author} {\bibfnamefont {P.~A.}\ \bibnamefont {Maksimov}}, \bibinfo
		{author} {\bibfnamefont {A.~L.}\ \bibnamefont {Chernyshev}}, \bibinfo
		{author} {\bibfnamefont {A.}~\bibnamefont {Honecker}}, \ and\ \bibinfo
		{author} {\bibfnamefont {R.}~\bibnamefont {Valent{\'\i}}},\ }\href@noop {}
	{\bibfield  {journal} {\bibinfo  {journal} {Nature communications}\ }\textbf
		{\bibinfo {volume} {8}},\ \bibinfo {pages} {1152} (\bibinfo {year}
		{2017}{\natexlab{b}})}\BibitemShut {NoStop}%
	\bibitem [{\citenamefont {Winter}\ \emph {et~al.}(2018)\citenamefont {Winter},
		\citenamefont {Riedl}, \citenamefont {Kaib}, \citenamefont {Coldea},\ and\
		\citenamefont {Valent{\'\i}}}]{winter2018probing}%
	\BibitemOpen
	\bibfield  {author} {\bibinfo {author} {\bibfnamefont {S.~M.}\ \bibnamefont
			{Winter}}, \bibinfo {author} {\bibfnamefont {K.}~\bibnamefont {Riedl}},
		\bibinfo {author} {\bibfnamefont {D.}~\bibnamefont {Kaib}}, \bibinfo {author}
		{\bibfnamefont {R.}~\bibnamefont {Coldea}}, \ and\ \bibinfo {author}
		{\bibfnamefont {R.}~\bibnamefont {Valent{\'\i}}},\ }\href@noop {} {\bibfield
		{journal} {\bibinfo  {journal} {Physical review letters}\ }\textbf {\bibinfo
			{volume} {120}},\ \bibinfo {pages} {077203} (\bibinfo {year}
		{2018})}\BibitemShut {NoStop}%
	\bibitem [{\citenamefont {Riedl}\ \emph {et~al.}(2022)\citenamefont {Riedl},
		\citenamefont {Amoroso}, \citenamefont {Backes}, \citenamefont {Razpopov},
		\citenamefont {Nguyen}, \citenamefont {Yamauchi}, \citenamefont {Barone},
		\citenamefont {Winter}, \citenamefont {Picozzi},\ and\ \citenamefont
		{Valent\'{\i}}}]{riedl2022microscopicorigin}%
	\BibitemOpen
	\bibfield  {author} {\bibinfo {author} {\bibfnamefont {K.}~\bibnamefont
			{Riedl}}, \bibinfo {author} {\bibfnamefont {D.}~\bibnamefont {Amoroso}},
		\bibinfo {author} {\bibfnamefont {S.}~\bibnamefont {Backes}}, \bibinfo
		{author} {\bibfnamefont {A.}~\bibnamefont {Razpopov}}, \bibinfo {author}
		{\bibfnamefont {T.~P.~T.}\ \bibnamefont {Nguyen}}, \bibinfo {author}
		{\bibfnamefont {K.}~\bibnamefont {Yamauchi}}, \bibinfo {author}
		{\bibfnamefont {P.}~\bibnamefont {Barone}}, \bibinfo {author} {\bibfnamefont
			{S.~M.}\ \bibnamefont {Winter}}, \bibinfo {author} {\bibfnamefont
			{S.}~\bibnamefont {Picozzi}}, \ and\ \bibinfo {author} {\bibfnamefont
			{R.}~\bibnamefont {Valent\'{\i}}},\ }\href {\doibase
		10.1103/PhysRevB.106.035156} {\bibfield  {journal} {\bibinfo  {journal}
			{Phys. Rev. B}\ }\textbf {\bibinfo {volume} {106}},\ \bibinfo {pages}
		{035156} (\bibinfo {year} {2022})}\BibitemShut {NoStop}%
	\bibitem [{\citenamefont {Razpopov}\ \emph {et~al.}(2023)\citenamefont
		{Razpopov}, \citenamefont {Kaib}, \citenamefont {Backes}, \citenamefont
		{Balents}, \citenamefont {Wilson}, \citenamefont {Ferrari}, \citenamefont
		{Riedl},\ and\ \citenamefont {Valent{\'\i}}}]{razpopov2023j}%
	\BibitemOpen
	\bibfield  {author} {\bibinfo {author} {\bibfnamefont {A.}~\bibnamefont
			{Razpopov}}, \bibinfo {author} {\bibfnamefont {D.~A.}\ \bibnamefont {Kaib}},
		\bibinfo {author} {\bibfnamefont {S.}~\bibnamefont {Backes}}, \bibinfo
		{author} {\bibfnamefont {L.}~\bibnamefont {Balents}}, \bibinfo {author}
		{\bibfnamefont {S.~D.}\ \bibnamefont {Wilson}}, \bibinfo {author}
		{\bibfnamefont {F.}~\bibnamefont {Ferrari}}, \bibinfo {author} {\bibfnamefont
			{K.}~\bibnamefont {Riedl}}, \ and\ \bibinfo {author} {\bibfnamefont
			{R.}~\bibnamefont {Valent{\'\i}}},\ }\href@noop {} {\bibfield  {journal}
		{\bibinfo  {journal} {npj Quantum Materials}\ }\textbf {\bibinfo {volume}
			{8}},\ \bibinfo {pages} {36} (\bibinfo {year} {2023})}\BibitemShut {NoStop}%
	\bibitem [{\citenamefont {Autieri}\ \emph {et~al.}(2022)\citenamefont
		{Autieri}, \citenamefont {Cuono}, \citenamefont {Noce}, \citenamefont
		{Rybak}, \citenamefont {Kotur}, \citenamefont {Agrapidis}, \citenamefont
		{Wohlfeld},\ and\ \citenamefont {Birowska}}]{autieri2022limited}%
	\BibitemOpen
	\bibfield  {author} {\bibinfo {author} {\bibfnamefont {C.}~\bibnamefont
			{Autieri}}, \bibinfo {author} {\bibfnamefont {G.}~\bibnamefont {Cuono}},
		\bibinfo {author} {\bibfnamefont {C.}~\bibnamefont {Noce}}, \bibinfo {author}
		{\bibfnamefont {M.}~\bibnamefont {Rybak}}, \bibinfo {author} {\bibfnamefont
			{K.~M.}\ \bibnamefont {Kotur}}, \bibinfo {author} {\bibfnamefont {C.~E.}\
			\bibnamefont {Agrapidis}}, \bibinfo {author} {\bibfnamefont {K.}~\bibnamefont
			{Wohlfeld}}, \ and\ \bibinfo {author} {\bibfnamefont {M.}~\bibnamefont
			{Birowska}},\ }\href@noop {} {\bibfield  {journal} {\bibinfo  {journal} {The
				Journal of Physical Chemistry C}\ }\textbf {\bibinfo {volume} {126}},\
		\bibinfo {pages} {6791} (\bibinfo {year} {2022})}\BibitemShut {NoStop}%
	\bibitem [{\citenamefont {Pavarini}(2014)}]{Pavarini_2014}%
	\BibitemOpen
	\bibfield  {author} {\bibinfo {author} {\bibfnamefont {E.}~\bibnamefont
			{Pavarini}},\ }\enquote {\bibinfo {title} {Electronic structure calculations
			with lda$+$dmft},}\ in\ \href {\doibase 10.1007/978-3-319-06379-9_18} {\emph
		{\bibinfo {booktitle} {Mathematical Physics Studies}}}\ (\bibinfo
	{publisher} {Springer International Publishing},\ \bibinfo {year} {2014})\
	p.\ \bibinfo {pages} {321–341}\BibitemShut {NoStop}%
	\bibitem [{\citenamefont {Jiang}\ \emph {et~al.}(2010)\citenamefont {Jiang},
		\citenamefont {Gomez-Abal}, \citenamefont {Rinke},\ and\ \citenamefont
		{Scheffler}}]{scheffler2010firstprinciplesmodeling}%
	\BibitemOpen
	\bibfield  {author} {\bibinfo {author} {\bibfnamefont {H.}~\bibnamefont
			{Jiang}}, \bibinfo {author} {\bibfnamefont {R.~I.}\ \bibnamefont
			{Gomez-Abal}}, \bibinfo {author} {\bibfnamefont {P.}~\bibnamefont {Rinke}}, \
		and\ \bibinfo {author} {\bibfnamefont {M.}~\bibnamefont {Scheffler}},\ }\href
	{\doibase 10.1103/PhysRevB.82.045108} {\bibfield  {journal} {\bibinfo
			{journal} {Phys. Rev. B}\ }\textbf {\bibinfo {volume} {82}},\ \bibinfo
		{pages} {045108} (\bibinfo {year} {2010})}\BibitemShut {NoStop}%
	\bibitem [{\citenamefont {Koepernik}\ and\ \citenamefont
		{Eschrig}(1999)}]{FPLO1999}%
	\BibitemOpen
	\bibfield  {author} {\bibinfo {author} {\bibfnamefont {K.}~\bibnamefont
			{Koepernik}}\ and\ \bibinfo {author} {\bibfnamefont {H.}~\bibnamefont
			{Eschrig}},\ }\href {\doibase 10.1103/PhysRevB.59.1743} {\bibfield  {journal}
		{\bibinfo  {journal} {Phys. Rev. B}\ }\textbf {\bibinfo {volume} {59}},\
		\bibinfo {pages} {1743} (\bibinfo {year} {1999})}\BibitemShut {NoStop}%
	\bibitem [{\citenamefont {Mou}\ \emph {et~al.}(2024)\citenamefont {Mou},
		\citenamefont {Zhang}, \citenamefont {Xiang}, \citenamefont {Xu},
		\citenamefont {Zhong}, \citenamefont {Cava}, \citenamefont {Zhou},
		\citenamefont {Jiang}, \citenamefont {Smirnov}, \citenamefont {Drichko} \emph
		{et~al.}}]{mou2024comparative}%
	\BibitemOpen
	\bibfield  {author} {\bibinfo {author} {\bibfnamefont {B.~S.}\ \bibnamefont
			{Mou}}, \bibinfo {author} {\bibfnamefont {X.}~\bibnamefont {Zhang}}, \bibinfo
		{author} {\bibfnamefont {L.}~\bibnamefont {Xiang}}, \bibinfo {author}
		{\bibfnamefont {Y.}~\bibnamefont {Xu}}, \bibinfo {author} {\bibfnamefont
			{R.}~\bibnamefont {Zhong}}, \bibinfo {author} {\bibfnamefont {R.~J.}\
			\bibnamefont {Cava}}, \bibinfo {author} {\bibfnamefont {H.}~\bibnamefont
			{Zhou}}, \bibinfo {author} {\bibfnamefont {Z.}~\bibnamefont {Jiang}},
		\bibinfo {author} {\bibfnamefont {D.}~\bibnamefont {Smirnov}}, \bibinfo
		{author} {\bibfnamefont {N.}~\bibnamefont {Drichko}},  \emph {et~al.},\
	}\href@noop {} {\bibfield  {journal} {\bibinfo  {journal} {arXiv preprint
				arXiv:2403.11980}\ } (\bibinfo {year} {2024})}\BibitemShut {NoStop}%
	\bibitem [{\citenamefont {Coury}\ \emph {et~al.}(2016)\citenamefont {Coury},
		\citenamefont {Dudarev}, \citenamefont {Foulkes}, \citenamefont {Horsfield},
		\citenamefont {Ma},\ and\ \citenamefont
		{Spencer}}]{spencer2016hubbardlikehamiltonians}%
	\BibitemOpen
	\bibfield  {author} {\bibinfo {author} {\bibfnamefont {M.~E.~A.}\
			\bibnamefont {Coury}}, \bibinfo {author} {\bibfnamefont {S.~L.}\ \bibnamefont
			{Dudarev}}, \bibinfo {author} {\bibfnamefont {W.~M.~C.}\ \bibnamefont
			{Foulkes}}, \bibinfo {author} {\bibfnamefont {A.~P.}\ \bibnamefont
			{Horsfield}}, \bibinfo {author} {\bibfnamefont {P.-W.}\ \bibnamefont {Ma}}, \
		and\ \bibinfo {author} {\bibfnamefont {J.~S.}\ \bibnamefont {Spencer}},\
	}\href {\doibase 10.1103/PhysRevB.93.075101} {\bibfield  {journal} {\bibinfo
			{journal} {Phys. Rev. B}\ }\textbf {\bibinfo {volume} {93}},\ \bibinfo
		{pages} {075101} (\bibinfo {year} {2016})}\BibitemShut {NoStop}%
	\bibitem [{\citenamefont {Tjeng}\ \emph {et~al.}(1990)\citenamefont {Tjeng},
		\citenamefont {Meinders}, \citenamefont {van Elp}, \citenamefont {Ghijsen},
		\citenamefont {Sawatzky},\ and\ \citenamefont
		{Johnson}}]{johnson1990electronicstructureofAg2O}%
	\BibitemOpen
	\bibfield  {author} {\bibinfo {author} {\bibfnamefont {L.~H.}\ \bibnamefont
			{Tjeng}}, \bibinfo {author} {\bibfnamefont {M.~B.~J.}\ \bibnamefont
			{Meinders}}, \bibinfo {author} {\bibfnamefont {J.}~\bibnamefont {van Elp}},
		\bibinfo {author} {\bibfnamefont {J.}~\bibnamefont {Ghijsen}}, \bibinfo
		{author} {\bibfnamefont {G.~A.}\ \bibnamefont {Sawatzky}}, \ and\ \bibinfo
		{author} {\bibfnamefont {R.~L.}\ \bibnamefont {Johnson}},\ }\href {\doibase
		10.1103/PhysRevB.41.3190} {\bibfield  {journal} {\bibinfo  {journal} {Phys.
				Rev. B}\ }\textbf {\bibinfo {volume} {41}},\ \bibinfo {pages} {3190}
		(\bibinfo {year} {1990})}\BibitemShut {NoStop}%
	\bibitem [{\citenamefont {Dhakal}\ \emph {et~al.}(2024)\citenamefont {Dhakal},
		\citenamefont {Griffith},\ and\ \citenamefont {Winter}}]{dhakal2024hybrid}%
	\BibitemOpen
	\bibfield  {author} {\bibinfo {author} {\bibfnamefont {R.}~\bibnamefont
			{Dhakal}}, \bibinfo {author} {\bibfnamefont {S.}~\bibnamefont {Griffith}}, \
		and\ \bibinfo {author} {\bibfnamefont {S.~M.}\ \bibnamefont {Winter}},\
	}\href@noop {} {\bibfield  {journal} {\bibinfo  {journal} {arXiv preprint
				arXiv:2403.05958}\ } (\bibinfo {year} {2024})}\BibitemShut {NoStop}%
	\bibitem [{\citenamefont {Dagotto}(1994)}]{dagotto1994correlated}%
	\BibitemOpen
	\bibfield  {author} {\bibinfo {author} {\bibfnamefont {E.}~\bibnamefont
			{Dagotto}},\ }\href {\doibase 10.1103/RevModPhys.66.763} {\bibfield
		{journal} {\bibinfo  {journal} {Rev. Mod. Phys.}\ }\textbf {\bibinfo {volume}
			{66}},\ \bibinfo {pages} {763} (\bibinfo {year} {1994})}\BibitemShut
	{NoStop}%
	\bibitem [{\citenamefont {van Elp}\ \emph {et~al.}(1991)\citenamefont {van
			Elp}, \citenamefont {Wieland}, \citenamefont {Eskes}, \citenamefont {Kuiper},
		\citenamefont {Sawatzky}, \citenamefont {de~Groot},\ and\ \citenamefont
		{Turner}}]{turner1991electronicLiCoO2}%
	\BibitemOpen
	\bibfield  {author} {\bibinfo {author} {\bibfnamefont {J.}~\bibnamefont {van
				Elp}}, \bibinfo {author} {\bibfnamefont {J.~L.}\ \bibnamefont {Wieland}},
		\bibinfo {author} {\bibfnamefont {H.}~\bibnamefont {Eskes}}, \bibinfo
		{author} {\bibfnamefont {P.}~\bibnamefont {Kuiper}}, \bibinfo {author}
		{\bibfnamefont {G.~A.}\ \bibnamefont {Sawatzky}}, \bibinfo {author}
		{\bibfnamefont {F.~M.~F.}\ \bibnamefont {de~Groot}}, \ and\ \bibinfo {author}
		{\bibfnamefont {T.~S.}\ \bibnamefont {Turner}},\ }\href {\doibase
		10.1103/PhysRevB.44.6090} {\bibfield  {journal} {\bibinfo  {journal} {Phys.
				Rev. B}\ }\textbf {\bibinfo {volume} {44}},\ \bibinfo {pages} {6090}
		(\bibinfo {year} {1991})}\BibitemShut {NoStop}%
\end{thebibliography}
\end{document}